\documentclass[10pt]{article}

\pdfoutput=1

\usepackage{ifpdf}
\usepackage[latin1]{inputenc}
\usepackage{graphicx}
\usepackage{graphics}
\usepackage{rotating}
\usepackage{amsmath}
\usepackage{amsthm}
\usepackage{amsfonts}
\usepackage{amssymb}
\usepackage{thmtools, thm-restate}
\usepackage{mathrsfs}
\usepackage{latexsym}
\usepackage{bbm} %Identity matrix \mathbbm{1} 
\usepackage{nicefrac}
\usepackage{enumerate}
\usepackage{url}
\usepackage{lscape}
\usepackage[section]{placeins} %keep floats in their sections
\usepackage[UKenglish]{babel}

\usepackage{tikz}
\usetikzlibrary{backgrounds,fit,decorations.pathreplacing,calc}

\DeclareMathOperator{\Tr}{Tr}
\DeclareMathOperator{\id}{id}

\newcommand{\bra}[1]{\left\langle #1 \right\rvert}
\newcommand{\ket}[1]{ \left\lvert #1\right\rangle}

\newcommand{\ketbra}[2]{\left\lvert #1 \rangle \langle #2 \right\rvert}

\newcommand{\norm}[1]{\left\lVert #1 \right\rVert}

\theoremstyle{plain}
\newtheorem{theorem}{Theorem}[section]
\newtheorem{corollary}[theorem]{Corollary}
\newtheorem{lemma}[theorem]{Lemma}
\newtheorem{proposition}[theorem]{Proposition}

\theoremstyle{definition}
\newtheorem{definition}{Definition}[section]

\theoremstyle{remark}

\numberwithin{equation}{section}

\begin{document}

\title{Universal Asymmetric Quantum Cloning Revisited}
\author{A.K.~Hashagen\\
\small Department of Mathematics,
\small Technische Universit\"at M\"unchen,
\small 85748 Garching,
\small Germany} 
\date{\small \today} 

\maketitle
\begin{abstract}
This paper revisits the universal asymmetric $1 \to 2$ quantum cloning problem. We identify the symmetry properties of this optimization problem, giving us access to the optimal quantum cloning map. Furthermore, we use the bipolar theorem, a famous method from convex analysis, to completely characterize the set of achievable single quantum clone qualities using the fidelity as our figure of merit; from this it is easier to give the optimal cloning map and to quantify the quality tradeoff in universal asymmetric quantum cloning. Additionally, it allows us to analytically specify  the set of achievable single quantum clone qualities using a range of different figures of merit. 
\end{abstract}
\tableofcontents

%%%%%%%%%%%%%%%%%%%%%%%%%%%%%%%%%%%%%%%%%%%%%
\section{Introduction}
%%%%%%%%%%%%%%%%%%%%%%%%%%%%%%%%%%%%%%%%%%%%%
\noindent
One of the most fundamental, but nevertheless intriguing, feature of quantum mechanics is the impossibility to perfectly clone an arbitrary quantum state. The intrinsic linearity of quantum mechanics facilitates this remarkable difference between classical information and its quantum counterpart, quantum information, which cannot be copied in a perfect manner. This observation is known as the ``No-Cloning Theorem~\ref{thm:nocloning}'' \cite{Wootters_Zurek_1982}. It is deeply intertwined with the impossibility of superluminal communication, the impossibility of classical teleportation  as well as the impossibility of fully determining an unknown quantum state \cite{Keyl_Habil_2003}. Even though the No-Cloning Theorem~\ref{thm:nocloning} gives rise to a lot of impossibilities, it also allows advantageous use within for example quantum cryptography. 

The possibilities that open up with the study of the No-Cloning Theorem~\ref{thm:nocloning} gave rise to a vast research area. This research was fuelled even further by experimental advancements; in these experiments approximate quantum cloning was realized using different techniques  \cite{lamas-linares_simon_howell_bouwmeester_2002, cummins_jones_furze_soffe_mosca_peach_jones_2002, zhao_zhang_zhou_chen_lu_karlsson_pan_2005, nagali_giovannini_marrucci_slussarenko_santamato_sciarrino_2010}.
One question, which turns out to be especially interesting is the question of approximate quantum cloning and its inherent boundaries. 
An intensive review is given by Cerf and Fiur\'{a}\v{s}ek \cite{cerf_fiurasek_2006} and by Scarani, Iblisdir and Gisin \cite{scarani_iblisdir_gisin_2005}. Even though perfect quantum cloning is impossible, it can be done in an approximate manner. This means that we are looking for a quantum channel, which clones any input state as good as possible. This is  called universal quantum cloning, because this setting is independent of the input state, i.e. the figure of merit assessing the quality of the clones is state-independent. 
In the case in which all clones have the same quality, the cloning procedure is named universal symmetric quantum cloning. If the clones may have different qualities, the cloning procedure is named universal asymmetric quantum cloning. The symmetric quantum cloning is thus a special case of the asymmetric quantum cloning.

Quantum cloning has been studied immensely after the universal symmetric $1 \to 2$ qubit quantum cloning machine was discovered by Bu\v{z}ek and Hillery in 1996  \cite{buzek_hillery_1996}\nocite{hillery_buzek_1997}.  Their machine was shown to be optimal by Bru\ss, DiVincenzo, Ekert, Fuchs, Macchiavello and Smolin two years later \cite{bruss_divincenzo_ekert_fuchs_macchiavello_smolin_1998}. At about the same time, Gisin and Massar presented their work on universal symmetric quantum cloning machines that transform $N$ identical qubits into $M$ identical clones and gave a numerical suggestion for optimality \cite{gisin_massar_1997}. Full optimality was then provided through an analytical proof by Bru\ss, Ekert and Macchiavello \cite{bruss_ekert_macchiavello_1998}. \nocite{hillery_buzek_1997}
Naturally, all these universal symmetric quantum cloning machines were extended to the qudit case. This was independently done by Bu\v{z}ek and Hillery \cite{Buzek_Hillery_1998} and Cerf \cite{cerf_1998}, who analyzed the $1 \to 2$ quantum cloning case for qudits, as well as Werner, who constructed the unique optimal symmetric $N \to M$ qudits quantum cloning machine and together with Keyl shows full optimality using group theoretical methods \cite{werner_1998, keyl_werner_1999}.  

The thorough exploration of the asymmetric case began with papers by Bu\v{z}ek, Hillery and Bednik \cite{buzek_hillery_bednik_1998}, Niu and Griffiths \cite{niu_griffiths_1998} and Cerf \cite{cerf_1998, cerf_2000},  in which they independently analyze and derive the universal asymmetric $1 \to 2$ qubit quantum cloning machine. Furthermore, they generalized their results from qubits to qudits \cite{braunstein_buzek_hillery_2001, cerf_2000_QI}.
The more general universal asymmetric $N \to M$ qubit cloning machines were introduced by Iblisdir, Ac\'{i}n, Gisin, Fiur\'{a}\v{z}ek, Filip and Cerf \cite{iblisdir_acin_cerf_filip_fiurasek_gisin_2005} and, using a technique from group theory, by Iblisdir, Ac\'{i}n and Gisin \cite{iblisdir_acin_gisin_2006}. The extensions to qudits was then presented by Fiur\'{a}\v{z}ek, Filip and Cerf in an additional paper \cite{fiurasek_filip_cerf_2005}. The inherent tradeoff among various output fidelities was further clarified and visualized by Jiang and Yu \cite{jiang_yu_2012}. Moreover, {\'C}wikli{\'n}ski, Horodecki and Studzi{\'n}ski further discussed the asymmetric quantum cloning case  in their paper \cite{cwiklinski_horodecki_studzinski_2012}, in which they provide a general result on an admissible region of fidelities for universal $1 \to N$ qubit quantum cloning machines. This result was extended to qudits by Studzi{\'n}ski, {\'C}wikli{\'n}ski, Horodecki and Mozrzymas using a general group representation approach \cite{studzinski_cwiklinski_horodecki_mozrzymas_2014}.  Simultaneously, Kay together with Ramanathan and Kaszlikowshi analyzed special cases of the universal $N \to M$ qudit quantum cloning machines, such as the universal asymmetric $1 \to N$ qudit cloning problem or the universal asymmetric $N-1 \to N$ qubit cloning problem \cite{kay_kaszlikowski_ramanathan_2009, kay_ramanathan_kaszlikowski_2013, kay_2014}.

This paper is concerned with the universal asymmetric $1 \to 2$ quantum cloning. We are thus interested in a quantum channel, also called optimal cloning map, that produces two good approximate clones from one input state, such that the qualities of these two clones must not be equal and are independent of the input state. In other words, if we fix the quality of one of the clones, the optimal cloning map maximizes the quality of the other clone independent of the input state. There exists a natural tradeoff between the qualities of the two clones: if the quality of one clone increases, intuitively it is clear that the quality of the other clone must decrease, complying to the No-Cloning Theorem~\ref{thm:nocloning}. The goal of this paper is to quantify this intuitive behavior regarding the quality of the two clones and to rediscover this optimal cloning map corresponding to universal asymmetric $1 \to 2$ quantum cloning. The arising asymmetric quantum cloning map agrees with previous results; we derive it, however, using methods from Eggeling and Werner \cite{eggeling_werner_2001} and Vollbrecht and Werner \cite{vollbrecht_werner_2001} originally used in order to study separability properties and entanglement measures under symmetry respectively. Furthermore, in this paper we analytically derive the set of achievable single quantum clone qualities using different figures of merit by means of convex analysis techniques. This powerful but simple method is what sets it apart from previous results. 

The paper is organized as follows: In the next chapter, we give a brief overview of the setting under consideration in this paper. Chapter~\ref{sec:FigureOfMerit} discusses figures of merit with which the quality of the clones are assessed. In order to quantify the asymmetric tradeoff in the quality of the clones, single clone figures of merit are investigated solely. In Chapter~\ref{sec:Twirling} we observe that the optimal quantum cloning channel is a quantum channel featuring specific symmetry properties. These symmetry properties determine the Choi-Jamiolkowski state. The Choi-Jamiolkowski channel state duality establishes that all properties of the quantum channel are encoded in the corresponding state. Reformulating the asymmetric quantum cloning problem using this Choi-Jamiolkowski state yields the optimal quantum cloning channel, given in Theorem~\ref{thm:OptimalCloning} in Chapter~\ref{sec:OptimalMap}. Furthermore, in this chapter, we draw the connection to semidefinite programming, which may also be used to solve the convex quantum cloning optimization problem. In Chapter~\ref{sec:Bipolar} we use the bipolar theorem, a technique known from convex analysis, to fully characterize the set of all attainable single quantum clone fidelities. Theorem~\ref{thm:OptimalCloningSet} summarizes this main result. Additionally, the set of all achievable single quantum clone qualities using a range of different figures of merit are given in Corollary~\ref{cor:QuantumCloningSet_all}. These sets are depicted in figures found in the appendix.

%%%%%%%%%%%%%%%%%%%%%%%%%%%%%%%%%%%%%%%%%%%%%
\section{The Setting of Universal Asymmetric Quantum Cloning}
\label{sec:Setting}
%%%%%%%%%%%%%%%%%%%%%%%%%%%%%%%%%%%%%%%%%%%%%
\noindent
We  consider systems on a finite dimensional Hilbert space $\mathcal{H}=\mathbb{C}^d$. Denote as $\mathcal{M}_d$ the set of all complex-valued $d \times d$-matrices. Every quantum state is described by a density matrix $\rho \in \mathcal{M}_d$ with normalization $\Tr\left[\rho \right] =1$ and positivity property $\rho \geq 0$. The set of all $d$-dimensional density matrices or quantum states is denoted as $\mathcal{D}_d := \left\{ \rho \in \mathcal{M}_d \middle\vert \rho \geq 0, \Tr \left[ \rho \right] = 1 \right\}$. A transformation of a quantum state is described by a quantum channel, which is a  completely positive trace preserving linear map $T: \mathcal{M}_d \to \mathcal{M}_{d'}$. Furthermore, we denote by $\mathcal{U}\left(d\right):= \left\{U \in \mathcal{M}_d \middle\vert UU^\ast = U^\ast U = \mathbbm{1} \right\}$ the unitary group acting on our Hilbert space $\mathcal{H}=\mathbb{C}^d$. Moreover, $\mathbbm{1}$ is the identity matrix in $\mathcal{M}_d$.

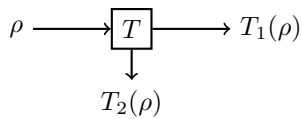
\begin{figure}[htp]
\vspace*{13pt}
	\centering
 \begin{tikzpicture}[thick]
    \tikzstyle{operator} = [draw,fill=white,minimum size=1.5em] 
    \tikzstyle{phase} = [draw,fill,shape=circle,minimum size=5pt,inner sep=0pt]
    \tikzstyle{surround} = [fill=blue!10,thick,draw=black,rounded corners=2mm]
    \matrix[row sep=0.4cm, column sep=0.8cm] (circuit) {
    % First row.
    \node (in) {$\rho $}; & 
    \node[operator] (T) {$T$}; &
		\node (out1) {$T_1(\rho)$}; \\
    % Second row.
		 & 
		\node (out2) {$T_2(\rho)$} ; &
		\\
    };
		\begin{pgfonlayer}{background}
        % Draw lines.
        \draw[thick, ->] (in) -- (T); 
				\draw[thick, ->] (T) -- (out1);
				\draw[thick, ->] (T) -- (out2);
    \end{pgfonlayer}
    \end{tikzpicture}
	\vspace*{13pt}
	\caption{\label{fig:Picture_Set_Up}The main setup of universal asymmetric $1 \to 2$ quantum cloning.}
\end{figure}

We are considering the universal asymmetric $1 \to 2$ quantum cloning case. The main setup is illustrated in Figure~\ref{fig:Picture_Set_Up}. The quantum cloning channel, $T \in \mathcal{B}(\mathcal{M}_d)$,
$T: \mathcal{M}_d \to \mathcal{M}_d \otimes \mathcal{M}_d $  is a trace preserving completely positive linear map with marginal maps 
$T_i: \mathcal{M}_d \to \mathcal{M}_d$ for $i=1,2$, defined as $T_i(\rho):=\Tr_{\bar{i}} \left[ T(\rho) \right]$, where the involution $i \mapsto \bar{i}$ corresponds to the permutation $\{ 2,1 \}$ of $\{1,2\}$\footnote{If $i=1$ then $\bar{i}=2$ and if $i=2$ then $\bar{i}=1$.}. Subscripts usually denote the underlying system. The corresponding Choi-matrix defined as $\mathcal{M}_{d^3} \ni \tau_{012} := \left( \id \otimes T \right) \left( \ketbra{\Omega}{\Omega} \right)$, where $\ketbra{\Omega}{\Omega}$ denotes the maximally entangled state, therefore has three subscripts, $1$ and $2$ corresponding to the two marginals of the quantum channel $T$ and a third subscript $0$ corresponding to the identity channel to which $T$ is tensored. 

Intuition lets us postulate, that the closer $T_1(\rho)$ is to $\rho$, the further away is $T_2(\rho)$ to $\rho$. Otherwise the No-Cloning Theorem~\ref{thm:nocloning} is violated. In order to analyze this intuition and to quantitatively describe it, some further definitions are needed. 

\noindent
\begin{theorem}[No-Cloning Theorem~\cite{Wootters_Zurek_1982}]
\label{thm:nocloning}
Consider quantum systems on a finite dimensional Hilbert space $\mathcal{H} = \mathbb{C}^d$.
There is no  completely positive trace preserving linear map, called a quantum channel, $T:\mathcal{M}_d \to \mathcal{M}_d \otimes \mathcal{M}_d$ such that for all quantum states $\rho \in \mathcal{D}_d$ the following holds, 
\begin{equation}
T(\rho) = \rho \otimes \rho.
\label{eq:nocloning}
\end{equation}
\end{theorem}

\noindent
\begin{proof}{
See \cite{Wootters_Zurek_1982}, or for the convenience of the reader we give a proof in the following.
The theorem is a consequence of linearity. Let $\{ \ketbra{\psi_i}{\psi_i} \}^n_{i=1}$ be a set of orthogonal pure states and $\{ \lambda_i \}^n_{i=1}$ be a set of probabilities such that $\lambda_i \neq 0$ for all $i$. If there was a map $T$ as specified in the theorem, then
\begin{equation}
\sum_i \lambda_i T(\ketbra{\psi_i}{\psi_i}) = \sum_i \lambda_i \ketbra{\psi_i}{\psi_i} \otimes \ketbra{\psi_i}{\psi_i}, 
\end{equation}
which has rank $n$, while,
\begin{equation}
\sum_i \lambda_i T(\ketbra{\psi_i}{\psi_i}) = T(\sum_i \lambda_i \ketbra{\psi_i}{\psi_i} ) = \sum_{ij} \lambda_i \lambda_j \ketbra{\psi_i}{\psi_i} \otimes \ketbra{\psi_j}{\psi_j}, 
\end{equation}
which has rank $n^2$}
\end{proof}

%%%%%%%%%%%%%%%%%%%%%%%%%%%%%%%%%%%%%%%%%%%%%
\section{The Figure of Merit assessing Single Clone Qualities}
\label{sec:FigureOfMerit}
%%%%%%%%%%%%%%%%%%%%%%%%%%%%%%%%%%%%%%%%%%%%%
\noindent
In order to assess the quality of the clones, we will consider a distance measure $d$ on the space of linear operators. This distance measure
$d(\cdot, \cdot): \mathcal{B}(\mathcal{M}_d) \times \mathcal{B}(\mathcal{M}_d) \to \mathbb{R}_+$ quantifies the quality of a clone. Since we are interested in the asymmetric tradeoff within the quality of the clones, the figure of merit is used to quantify the quality of a single clone. We will thus compare each marginal $T_i$ to the identity map; that is, we are going to consider $d(T_i, \id)$ for $i=1,2$. Our goal is to fully specify the set of all attainable single quantum clone qualities,
\[
\mathcal{C} = \left\{ z \in \mathbb{R}^2 \middle\vert z = \begin{pmatrix} d(T_1, \id) \\ d(T_2, \id) \end{pmatrix} \right\},
\]
using this figure of merit. 

\vspace*{12pt}
\noindent
\textbf{Required properties of our figure of merit:} Let $L,S:\mathcal{M}_d \to \mathcal{M}_d$ be quantum channels. We require the figure of merit to have the following properties due to technical reasons.
\begin{enumerate}[(i)]
\item Joint concavity: \[ d\left(L, S\right) \geq \lambda d\left( L^{(1)}, S^{(1)}\right)+(1-\lambda)d\left(L^{(2)}, S^{(2)}\right)\] for all $L$ and $S$, where  $L=\lambda L^{(1)}+(1-\lambda)L^{(2)}$ and  $S=\lambda S^{(1)}+(1-\lambda)S^{(2)}$, with $\lambda \in [0,1]$.\\
\item Unitary invariance: \[d (\mathbf{U} \circ L \circ \mathbf{U}^\ast, \mathbf{U} \circ S \circ \mathbf{U}^\ast)= d(L, S)\] for all  ideal channels $\mathbf{U}$ defined by $\mathbf{U}(\rho) = U\rho U^\ast$ with unitary $U \in \mathcal{U}\left(d\right)$ and where $\cdot^\ast$ denotes the adjoint or conjugate transpose. 
\item Furthermore, for reasons that will become clear later, we require that the origin is attainable, i.e. that $\{ 0 \} \in \mathcal{C}$. This requirement means that we are not necessarily considering a metric as a distance measure.
\end{enumerate}

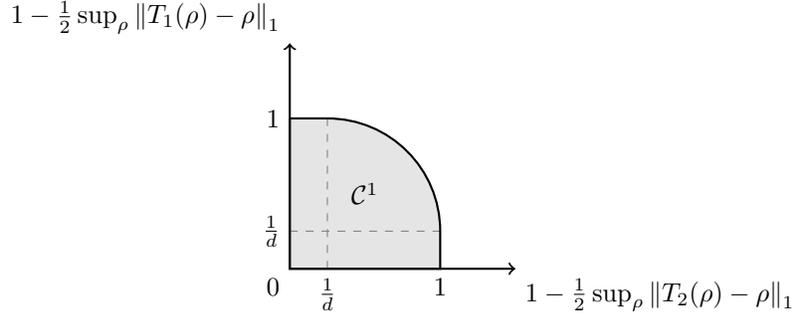
\begin{figure}[htb]
\vspace*{13pt}
	\centering
    \begin{tikzpicture}[thick]
		\draw [<->] (0,3) -- (0,0) -- (3,0);
		\draw [thin, gray, dashed] (0.5,0) -- (0.5,2);
		\draw [thin, gray, dashed] (0,0.5) -- (2,0.5);
		\draw [fill = gray, fill opacity=0.2] (0,2) -- (0.5,2) to [out=0, in = 90] (2,0.5) -- (2,0) -- (0,0) -- (0,2);
		\node [left] at (0,2) {$1$};
		\node [below] at (2,0) {$1$};
		\node [left] at (0,0.5) {$\frac{1}{d}$};
		\node [below] at (0.5,0) {$\frac{1}{d}$};
		\node [below left] at (0,0) {$0$};
		\node [above left] at (0,3) {$1 - \frac{1}{2}\sup_\rho \norm{T_1(\rho) - \rho}_1 $};
  	\node [below right] at (3,0) {$1 - \frac{1}{2}\sup_\rho \norm{T_2(\rho) - \rho}_1 $};
		\node at (1,1) {$\mathcal{C}^1$};
		\end{tikzpicture}
		\vspace*{13pt}
	\caption{\label{fig:Picture_Set_Trace_Norm}Set of all attainable single quantum clone qualities $\mathcal{C}^1$. The figure of merit is $d^1(T_i, \id) =  1- \frac{1}{2}\sup_\rho \norm{T_i(\rho)-\rho}_1$ for $i=1,2$.}
\end{figure}

An example of a valid distance measure that fulfills these properties is given by a variant of the induced trace norm distance 
\begin{equation}
d^1(T_i, \id) = 1- \frac{1}{2}\sup_{\rho \in \mathcal{D}_d} \norm{T_i(\rho)-\rho}_1.
\end{equation} 
It characterizes the maximum probability of not distinguishing the outputs of the two channels $T_i$, $i=1,2$,  and the identity channel $\id$ over all pure state inputs. 
This distance measure is specifically chosen in this way to always contain the origin, 
\[
\{0\} \in \mathcal{C}^1 = \left\{ z \in \mathbb{R}^2 \middle\vert z = \begin{pmatrix} d^1(T_1, \id) \\ d^1(T_2, \id) \end{pmatrix} \right\},
\] 
as illustrated in Figure~\ref{fig:Picture_Set_Trace_Norm}, for reasons that will become clear later. Furthermore, we can notice that if $1- \frac{1}{2}\sup_{\rho} \norm{T_i(\rho)-\rho}_1 = 1$, then the marginal must be given by $T_i(\rho)=\rho$, with $i=1,2$. 
If we now let the other marginal be given by $T_{\bar{i}}(\rho)=\sigma$, with $\bar{i}=2,1$, for some fixed quantum state $\sigma \in \mathcal{D}_d$, then  $1- \frac{1}{2}\sup_{\rho} \norm{T_{\bar{i}}(\rho)-\rho}_1 = \lambda_\text{min}(\sigma)$, where $\lambda_\text{min}(\sigma) \in \left[ 0, \frac{1}{d}\right] $ is the smallest eigenvalue of $\sigma$. These boundary points are visualized in Figure~\ref{fig:Picture_Set_Trace_Norm}.

Another example, which is going to be of interest to us later on, is the fidelity 
\begin{equation}
d^F(T_i, \id) = \bra{\Omega} \tau_{0i}\ket{\Omega},
\end{equation}
where $\tau_{0i} := \id \otimes T_i \left(\ketbra{\Omega}{\Omega}\right)$ is the Choi-Jamiolkowski state of the marginal map and $\ketbra{\Omega}{\Omega}$ is the maximally entangled state with $\ket{\Omega} = \frac{1}{\sqrt{d}}\sum_{i=1}^d \ket{ii}$. It measures the overlap of the output with the maximally entangled state, if $T_i$ acts on half a maximally entangled state.  %The set of all attainable single quantum clone qualities using the fidelity as figure of merit is illustrated in Figure \ref{fig:Picture_Set_Fidelity}.

Other examples that might be of interest are a variant of the induced Frobenius norm distance
\begin{equation}
d^2(T_i, \id) = 1- \sup_{\rho  \in \mathcal{D}_d} \norm{T_i(\rho)-\rho}_2,
\end{equation}
a variant of the induced operator norm distance
\begin{equation}
d^\infty(T_i, \id) = 1- \sup_{\rho  \in \mathcal{D}_d} \norm{T_i(\rho)-\rho}_\infty,
\end{equation}
and a variant of the diamond norm distance, which is a stabilized version of the induced trace norm distance,
\begin{equation}
d^\diamond(T_i, \id) = 1- \frac{1}{2}\norm{T_i-\id}_\diamond = 1 - \frac{1}{2}\sup_{\rho  \in \mathcal{D}_{d^2}} \norm{\left( T_i \otimes \id_d \right)(\rho)-\rho}_1,
\end{equation}
with $i=1,2$.
Note that all these distance measures are adjusted by $d^k_\text{max}$, $k=1,2,\infty, \diamond$, the maximum value that the norm may take such that the origin is always contained in the corresponding set of all attainable single quantum clone qualities. This is the case, because we always look at the worst case scenario over all quantum states.

The goal is to characterize all possible quantum clone qualities; it is thus of interest to us to consider the following optimization problem
\begin{equation}
\sup_T \left[ z_1 d^k(T_1, \id) + z_2 d^k(T_2, \id) \right],
\label{eq:supremum1}
\end{equation} 
with $z_1,z_2\in \mathbb{R}$ for different figures of merit, $k=F, 1,2,\infty, \diamond$. This optimization problem gives the upper boundary of the set of all attainable single quantum clone qualities.  
The set of admissible quantum channels $T$ is compact, since the set is bounded and closed in a finite dimensional vector space. The supremum in Eq.~(\ref{eq:supremum1}) is therefore attained for some optimal quantum channel $T_\text{optimal}$, which is the optimal quantum cloning channel, as it gives the best tradeoff possible within the qualities of the quantum clones. 

%%%%%%%%%%%%%%%%%%%%%%%%%%%%%%%%%%%%%%%%%%%%%
\section{The Symmetry Properties of the Optimal Quantum Cloning Channel}
\label{sec:Twirling}
%%%%%%%%%%%%%%%%%%%%%%%%%%%%%%%%%%%%%%%%%%%%%
\noindent
In order to find the optimal quantum cloning channel, it is of interest to us to identify special symmetry properties of this quantum channel. Let us define what we mean by a symmetrized quantum channel, because it turns out that the optimal quantum cloning channel is exactly of this type. 

\noindent
\begin{definition}[Symmetrized quantum channel]
\label{def:symmetrisedT}
A symmetrized quantum channel $\tilde{T}:\mathcal{M}_d \to \mathcal{M}_d$ is defined via the map 
\begin{align}
T\left(\cdot\right) \mapsto \tilde{T}\left(\cdot\right) &= \int_{\mathcal{U}\left(d\right)}\mathbf{U} \circ T \circ \mathbf{U}^\ast \left( \cdot \right) \,\mathrm{d}\mathbf{U} \nonumber \\
&= \int_{\mathcal{U}\left(d\right)}UT\left(U^\ast \cdot U\right) U^\ast  \,\mathrm{d}U \ \ \forall \ U \in \mathcal{U}\left(d\right), 
\end{align}
where $\mathrm{d}U$ denotes the normalized Haar measure on the unitary group $\mathcal{U}\left(d\right)$ and $\mathbf{U}\left(\cdot\right) = U\cdot U^\ast$ is the ideal quantum channel. Note that we will always use a tilde to denote a symmetrized quantum channel.
\end{definition}
This symmetrization (also called twirling and in quantum information first introduced in \cite{eggeling_werner_2001, vollbrecht_werner_2001}) can be considered as averaging over the unitary group $\mathcal{U}\left(d\right)$ on our Hilbert space $\mathcal{H}$. Let us consider a symmetrized quantum cloning channel $\tilde{T}:\mathcal{M}_d \to \mathcal{M}_d \otimes \mathcal{M}_d$ with
\[T(\rho) \mapsto \tilde{T}(\rho) = \int_{\mathcal{U}\left(d\right)}  (U \otimes U) T(U^\ast \rho U)(U \otimes U)^\ast \,\mathrm{d}U\] for every quantum state $\rho \in \mathcal{D}_d$. It turns out that this symmetry property also applies to its marginals; see the following Lemma~\ref{lem:MarginalsSymm}.

\noindent
\begin{lemma}
\label{lem:MarginalsSymm}
The marginal maps $T_i: \mathcal{M}_d \to \mathcal{M}_d$, $i=1,2$, of a symmetrized quantum channel $\tilde{T}: \mathcal{M}_d \to \mathcal{M}_d \otimes \mathcal{M}_d$ are symmetrized,
\[\tilde{T}_i (\rho) = \int_{\mathcal{U}\left(d\right)} U T_i \left( U^\ast \rho U \right)  U^\ast \, \mathrm{d} U,\] 
for $i=1,2$, for every quantum state $\rho \in \mathcal{D}_d$.
\end{lemma}

\noindent
\begin{proof}{
The marginal maps of a symmetrized quantum channel are given by
\begin{align*}
  \tilde{T}_{i} (\rho) &= \Tr_{\bar{i}} \left[ \tilde{T}(\rho) \right] \\
	&= \Tr_{\bar{i}} \left[ \int_{\mathcal{U}\left(d\right)} \left( U \otimes U \right) T \left( U^\ast \rho U \right) \left( U \otimes U \right)^\ast \, \mathrm{d} U  \right] \\
	&= \int_{\mathcal{U}\left(d\right)} \Tr_{\bar{i}} \left[  \left( U \otimes U \right) T \left( U^\ast \rho U \right) \left( U \otimes U \right)^\ast  \right] \, \mathrm{d} U  \\
	&= \int_{\mathcal{U}\left(d\right)} U \Tr_{\bar{i}} \left[   T \left( U^\ast \rho U \right)  \right]U^\ast \, \mathrm{d} U  \\
 &=\int_{\mathcal{U}\left(d\right)} U   T_{i} \left( U^\ast \rho U \right)  U^\ast \, \mathrm{d} U,
	\end{align*}
for both marginals $i=1,2$}
\end{proof}
These symmetrized marginal maps satisfy the so-called covariance property $\tilde{T}_i(V \rho V^\ast) = V \tilde{T}_i (\rho) V^\ast$ with $i=1,2$ for unitary $V \in \mathcal{U}\left(d\right)$, as stated in the Lemma~\ref{lem:Covariant} below \cite{eggeling_werner_2001, vollbrecht_werner_2001}.

\noindent
\begin{definition}[Covariant]
A quantum channel  $T:\mathcal{M}_d \to \mathcal{M}_d$ is called covariant with respect to $V$ if 
\[
T\left( V \cdot V^\ast \right) = V T\left( \cdot \right) V^\ast 
\]
holds for all $V \in \mathcal{U}\left(d\right)$.
\end{definition}

\noindent
\begin{lemma}
\label{lem:Covariant}
A symmetrized quantum channel $\tilde{T}: \mathcal{M}_d \to \mathcal{M}_d$ is covariant.
\end{lemma}

\noindent
\begin{proof}{
Using the definition of a symmetrized quantum channel yields
\begin{align*}
& \tilde{T}\left( V \cdot V^\ast\right) \\
=& \int_{\mathcal{U}\left(d\right)}{U T\left(U^\ast V \cdot V^\ast U \right) U^\ast \,\mathrm{d}U} \\
=& V \int_{\mathcal{U}\left(d\right)}{V^\ast U T\left(U^\ast V \cdot V^\ast U \right) U^\ast V \,\mathrm{d}U} V^\ast  \\
=& V \int_{\mathcal{U}\left(d\right)}{W T\left(W^\ast \cdot W \right) W^\ast  \,\mathrm{d}(VW)} V^\ast  \\
=& V \int_{\mathcal{U}\left(d\right)}{W T\left(W^\ast \cdot W \right) W^\ast  \,\mathrm{d}(W)} V^\ast \\
=& V \tilde{T}(\cdot) V^\ast,
\end{align*}
where we have defined $W:=V^\ast U$ with unitaries $U, V \in \mathcal{U}\left(d\right)$ and used the properties of the Haar measure}
\end{proof}

The figure of merit that assesses the single clone qualities is influenced by this covariance property in the following way, namely
\begin{align*}
d^k (\tilde{T}_i, \id) &= d^k \left( \int_{\mathcal{U}\left(d\right)} \mathbf{U} \circ T_i \circ \mathbf{U}^\ast \, \mathrm{d}\mathbf{U} , \id \right) \\
&\geq  \int_{\mathcal{U}\left(d\right)} d^k \left(  \mathbf{U} \circ T_i \circ \mathbf{U}^\ast, \id \right) \, \mathrm{d}\mathbf{U}  \\
&= \int_{\mathcal{U}\left(d\right)} d^k \left(  T_i, \mathbf{U} \circ \id  \circ \mathbf{U}^\ast \right)  \, \mathrm{d}\mathbf{U}  \\
&= d^k(T_i, \id),
\end{align*}
for $k=F, 1,2,\infty, \diamond$, where we have used the joint concavity property (i) of the figure of merit.
The optimization problem given by Eq.~(\ref{eq:supremum1}) therefore simplifies, because the supremum is attained for a symmetrized quantum channel $\tilde{T}$, i.e. we have
\begin{equation}
\sup_T \left[ z_1 d^k(T_1, \id) + z_2 d^k(T_2, \id) \right] = \sup_{\tilde{T}} \left[ z_1 d^k(\tilde{T}_1, \id) + z_2 d^k(\tilde{T}_2, \id) \right],
\label{eq:supremum2}
\end{equation}
with $z_1,z_2 \in \mathbb{R}$, for $k=F, 1,2,\infty, \diamond$. 

%%%%%%%%%%%%%%%%%%%%%%%%%%%%%%%%%%%%%%%%%%%%%
\section{The Optimal Quantum Cloning Channel}
\label{sec:OptimalMap}
%%%%%%%%%%%%%%%%%%%%%%%%%%%%%%%%%%%%%%%%%%%%%
\noindent
The last chapter has shown that the optimal quantum cloning channel is of a symmetrized form. This gives rise to a specific structure of its Choi-Jamiolkowski state, which we are going to exploit to solve the optimization problem in Eq.~(\ref{eq:supremum2}). This chapter therefore discusses the implication of the symmetrized optimal quantum cloning channel on its Choi-Jamiolkowski state and uses this additional structure to derive the optimal quantum cloning channel.

The Choi-Jamiolkowski state of a quantum channel $T$ is defined as
\begin{equation}
\tau := \id \otimes T \left( \ketbra{\Omega}{\Omega} \right),
\end{equation}
where $\ket{\Omega}= \frac{1}{\sqrt{d}} \sum_{i=1}^d \ket{ii}$.
We would like to simplify our optimization problem given by Eq.~(\ref{eq:supremum2}) even further using this Choi-Jamiolkowski state. For this purpose, we would like to show that $\left[ \tau, \bar{U} \otimes U \otimes U\right] =0$ for unitary $U \in \mathcal{U}\left(d\right)$, in the case of a symmetrized quantum channel, where $\bar{\cdot}$ denotes the complex conjugate.

\noindent
\begin{restatable}{lemma}{TwirlComm}
\label{lem:equiv}
For a Choi-Jamiolkowski state $\tau \in \mathcal{M}_{d^3}$,
\[
\left[ \tau, \bar{U} \otimes U \otimes U \right] = 0 \ \ \forall \ U \in \mathcal{U}\left(d\right) 
\]
is equivalent to
\[
\int_{\mathcal{U}\left(d\right)} \left( \bar{U} \otimes U \otimes U \right) \tau \left( \bar{U} \otimes U \otimes U \right)^\ast \,\mathrm{d} U = \tau.
\]
\end{restatable}
The proof of this Lemma~\ref{lem:equiv} can be found in the Appendix~\ref{app:lemmaequiv}.
We may now use this Lemma~\ref{lem:equiv} to prove the following Corollary~\ref{cor:ChoiStateofSymmT}.

\noindent
\begin{corollary} 
\label{cor:ChoiStateofSymmT}
Let $\tau := \id \otimes \tilde{T} \left( \ketbra{\Omega}{\Omega} \right)$ be the Choi-Jamiolkowski state of some symmetrized quantum channel $\tilde{T}$,  as in Definition~\ref{def:symmetrisedT}, then the following holds
\[
\left[ \tau , \bar{U}\otimes U \otimes U \right] = 0 \ \ \forall \ U \in \mathcal{U}\left(d\right).
\]
\end{corollary}

\noindent
\begin{proof}{
Remembering that
$
U\otimes \mathbbm{1} \ket{\Omega} = \mathbbm{1} \otimes U^T \ket{\Omega},
$
such that
$
U \otimes \bar{U} \ket{\Omega} = \ket{\Omega},
$
we get
\begin{align*}
&\int_{\mathcal{U}\left(d\right)} \left( \bar{U} \otimes U \otimes U \right) \tau \left( \bar{U} \otimes U \otimes U \right)^\ast  \,\mathrm{d} U \\
= &\int_{\mathcal{U}\left(d\right)} \left( \bar{U} \otimes U \otimes U \right) \left( \id \otimes \tilde{T} \ketbra{\Omega}{\Omega} \right) \left( \bar{U} \otimes U \otimes U \right)^\ast  \,\mathrm{d} U \\ 
= & \int_{\mathcal{U}\left(d\right)} \left( \mathbbm{1} \otimes U \otimes U \right) \left( \id \otimes \tilde{T} \right) \left( \left( \mathbbm{1} \otimes U^\ast \right) \ketbra{\Omega}{\Omega} \left( \mathbbm{1} \otimes U \right) \right) \left( \mathbbm{1} \otimes U \otimes U \right)^\ast  \,\mathrm{d} U \\
= & \tau,
\end{align*} 
due to the covariance property of $\tilde{T}$. Application of Lemma~\ref{lem:equiv} finishes the proof}
\end{proof}

\noindent
\begin{restatable}{proposition}{UUU}
\label{prop:UUU}
If for a Choi-Jamiolkowski state $\tau \in \mathcal{M}_{d^3}$ we have that
\[
\left[ \tau, \bar{U} \otimes U \otimes U \right] = 0 \ \ \forall \  U \in \mathcal{U}\left(d\right),
\]
then
\[
\left[ \tau^{t_0}, U \otimes U \otimes U \right] = 0,
\]
where $t_0$ denotes the partial transpose on the first system. 
\end{restatable}
The proof of this Proposition~\ref{prop:UUU} can be found in the Appendix~\ref{app:propUUU}.

The optimization problem given by Eq.~(\ref{eq:supremum2}) was reduced to a supremum over all symmetrized quantum channels, because we found that the optimal quantum cloning channel must be of this form. Using Corollary~\ref{cor:ChoiStateofSymmT}, we may, without loss of generality, restrict to quantum cloning channels whose Choi-Jamiolkowski matrix $\tau$ commutes with $\left\{ \bar{U} \otimes U \otimes U: U \in \mathcal{U}\left(d\right) \right\}$ and we would like to reformulate our problem by means of $\tau$. 

\noindent
\begin{theorem}[Weyl {\cite[Chapter IV]{weyl_1997_classical_groups}}]
\label{thm:weyl}
Let $\mathcal{H}$ be a finite-dimensional Hilbert space. If an operator $\tau$ acting on $\mathcal{H}^{\otimes n}$ fulfills $\left[ \tau, U^{\otimes n} \right] = 0$ for all unitaries $U \in \mathcal{U}\left(d\right)$, then it is a linear combination of operators $V_\pi$ representing the permutation group on $\mathcal{H}^{\otimes n}$,
\[
\tau = \sum_{\pi \in \mathrm{S}_n} a_\pi V_\pi,
\]
where S$_n$ is the symmetric group on $n$ elements, $\pi$ are all possible permutations of $n$ elements and $a_\pi \in \mathbb{C}$. The permutation operators $V_\pi$ are defined via
\[
V_\pi \left( v_1 \otimes \ldots \otimes v_n \right)= v_{\pi^{-1}(1)} \otimes \ldots \otimes v_{\pi^{-1}(n)}.
\]
\end{theorem}

\noindent
\begin{proof}{
The theorem immediately follows from \cite[Theorem IX.11.5]{simon_representations}. Denote by SU$(d)$ the special unitary group of finite degree $d$ and by S$_n$ the symmetric group on $n$ elements. Let $\mathcal{A}$ be the group algebra of SU$(d)$  and $\mathcal{B}$ be the group algebra of S$_n$ generated by their unitary representation on $\mathcal{H}$. Since SU$(d)$ and S$_n$ act dually on $\left(\mathbb{C}^d\right)^{\otimes n}$, we have $\mathcal{A}'=\mathcal{B}$. The commutant is thus exactly the algebra generated by the permutation operators $V_\pi$. If an operator commutes with all unitaries of the form $U^{\otimes n}$, it must therefore be an element of this algebra, i.e. a linear combination of permutation operators} 
\end{proof}

Considering Lemma~\ref{prop:UUU}, we know that $\tau^{t_0}$ commutes with all unitaries of the form $U\otimes U \otimes U$. Furthermore, by Theorem~\ref{thm:weyl},  $\tau^{t_0}$ must be given by a linear combination of permutation operators, in our case acting on three elements, 
\begin{equation}
\tau_{012}^{t_0} = \sum_{\pi \in \mathrm{S}_3} a_\pi V_\pi = a_1 \mathbbm{1} + a_2 V_{(01)} + a_3 V_{(02)} + a_4 V_{(12)} + a_5 V_{(012)} + a_6 V_{(210)},
\end{equation}
with $a_\pi \in \mathbb{C}$, where $V_{(01)}$ denotes the permutation operator of the first two factors (similarly for $V_{(02)}$ and $V_{(12)}$), $V_{(012)}$ denotes the cyclic permutation and similarly $V_{(210)}$ denotes the anticyclic permutation \cite{eggeling_werner_2001, vollbrecht_werner_2001}. 

The marginal maps are thus given as
\begin{subequations}
\begin{align}
\tau_{01}^{t_0} &= \Tr_2 [\tau_{012}^{t_0}] = (a_1d+a_3+a_4) \mathbbm{1} + (a_2d+a_5+a_6) \mathbb{F}, \\
\tau_{01} &= (a_1d+a_3+a_4) \mathbbm{1} + (a_2d+a_5+a_6) d \ketbra{\Omega}{\Omega}, \\
\tilde{T}_1(\rho) &= (a_1d+a_3+a_4) d \mathbbm{1} \Tr[\rho] + (a_2d+a_5+a_6) d \rho, 
\end{align}
\end{subequations}
and
\begin{subequations}
\begin{align}
\tau_{02}^{t_0} &=  \Tr_1 [\tau_{012}^{t_0}] = (a_1d+a_2+a_4) \mathbbm{1} + (a_3d+a_5+a_6) \mathbb{F}, \\
\tau_{02} &=   (a_1d+a_2+a_4) \mathbbm{1} + (a_3d+a_5+a_6)  d \ketbra{\Omega}{\Omega}, \\
\tilde{T}_2(\rho) &= (a_1d+a_2+a_4) d \mathbbm{1} \Tr[\rho] + (a_3d+a_5+a_6) d \rho,
\end{align}
\end{subequations}
with $a_1, \ldots, a_6 \in \mathbb{C}$, where we again denote by $\ketbra{\Omega}{\Omega} := \frac{1}{d} \sum_{i,j=1}^{d} \ketbra{ii}{jj}$ the maximally entangled state and by $\mathbb{F}:= \sum_{i,j=1}^d \ketbra{ji}{ij}$ the flip (or swap) operator.

As a quantum channel, $\tilde{T}$ is a completely positive trace preserving linear map and it must thus fulfill specific properties.  Due to the Choi-Jamiolkowski state-channel duality, the operator $\tau$ encodes all of its properties \cite[Chapter 4.4.3]{heinosaari_ziman_2012}. Denote by  $\tilde{T}^\ast$ the dual of the quantum channel $\tilde{T}$ corresponding to the Heisenberg picture.  \\
\noindent \textbf{Properties:} 
\begin{enumerate}[(i)]
	\item Hermiticity: $\tau = \tau^\ast$, i.e. \[a_1,\ldots, a_4 \in \mathbb{R} \text{ and } a_5 = \bar{a}_6 \in \mathbb{C}.\]
	\item Normalization: $\Tr [ \tau ] = \frac{1}{d} \Tr \left[ \tilde{T}^\ast(\mathbbm{1}) \right]$, i.e. \[a_1d^3+ (a_2+a_3+a_4)d^2+(a_5+a_6)d=1.\]
	\item Preservation of trace: $\tilde{T}^\ast(\mathbbm{1}) = \mathbbm{1}$ if and only if $\Tr_{12} \left[ \tau \right] = \frac{\mathbbm{1}}{d}$.
	\item Complete positivity: $\tilde{T}$  is completely positive if and only if $\tau \geq 0$.
\end{enumerate}
Note that if $\tilde{T}(\rho)$ is a completely positive trace preserving linear map, then so are its marginal maps $\tilde{T}_i(\rho):= \Tr_{\bar{i}}\left[ \tilde{T}(\rho)\right]$, $i=1,2$. 

In order to simplify notation and to visualize agreement to previously known results \cite{braunstein_buzek_hillery_2001, cerf_2000_QI}, let
\begin{align*}
(\alpha_1)^2 &:= a_1d^3+a_3d^2+a_4d^2, & (\beta_1)^2 &:= a_2d^2+a_5d+a_6d, \\
(\alpha_2)^2 &:= a_1d^3+a_2d^2+a_4d^2, & (\beta_2)^2 &:= a_3d^2+a_5d+a_6d.
\end{align*}
Then the Choi-Jamiolkowski states $\tau_{0i}$, $i=1,2$,  of the marginal maps $\tilde{T}_i$ are 
\begin{equation} 
\tau_{0i} = \alpha_i^2 \frac{\mathbbm{1}}{d^2} + \beta_i^2 \ketbra{\Omega}{\Omega}.
\end{equation}
The preservation of trace, property (iii), namely $\Tr_i \left[ \tau_{0i} \right] = \nicefrac{\mathbbm{1}}{d}$, gives a condition on $\beta_i$, namely that
\begin{align*}
\Tr_i \left[ \tau_{0i} \right] = \Tr_i \left[ \alpha_i^2 \frac{\mathbbm{1}}{d^2} + \beta_i^2 \ketbra{\Omega}{\Omega}\right] = (\alpha_i^2 +\beta_i^2)\frac{\mathbbm{1}}{d} = \frac{\mathbbm{1}}{d} \\
\Leftrightarrow \beta_i^2 = 1 - \alpha_i^2.
\end{align*}
Another property that the marginals must fulfill is complete positivity, property (iv), namely $\tau_{0i} \geq 0$. This yields
\begin{align*}
\tau_{0i} = \alpha_i^2 \frac{\mathbbm{1}}{d^2} + \beta_i^2 \ketbra{\Omega}{\Omega} \geq 0 \\
\Leftrightarrow \alpha_i^2 \geq 0 \text{ and } \beta_i^2 \geq - \frac{\alpha_i^2}{d^2}.
\end{align*}
Therefore, the marginal maps and their corresponding Choi-Jamiolkowski states are given as
\begin{subequations}
\begin{align}
\tau_{0i} &= \alpha_i^2 \frac{\mathbbm{1}}{d^2} + (1-\alpha_i^2) \ketbra{\Omega}{\Omega}, \\
\tilde{T}(\rho) &= \alpha_i^2 \frac{\mathbbm{1}}{d} \Tr[\rho] +  (1-\alpha_i^2) \rho,
\end{align}
\end{subequations}
with $\alpha_i^2 \in \left[ 0, \frac{d^2}{d^2-1}\right]$.

Since these properties must not only hold for the marginals, but also for the full quantum channel, consider
\begin{equation}
\tau_{012} = \sum_{\pi \in \mathrm{S}_3} a_\pi V_\pi^{t_0}.
\end{equation}
The preservation of trace property (iii), namely that $\Tr_{12} \left[ \tau_{012} \right] = \nicefrac{\mathbbm{1}}{d}$, yields
\begin{align*}
\Tr_{12} \left[ \tau_{012} \right] = \Tr_{12} \left[ \sum_{\pi \in \mathrm{S}_3} a_\pi V_\pi^{t_0} \right] = (a_1d^2+a_2d+a_3d+a_4d+a_5+a_6)\mathbbm{1} = \frac{\mathbbm{1}}{d} \\
\Leftrightarrow a_1d^2+a_2d+a_3d+a_4d+a_5+a_6 = \frac{1}{d}.
\end{align*}
Deciding complete positivity, property (iv), is a bit more tricky and we thus follow the idea of the following papers \cite{eggeling_werner_2001, vollbrecht_werner_2001}. One should first of all notice that \[\mathcal{A}^{t_0} := \left\{ \sum_{\pi \in \mathrm{S}_3} a_\pi V_\pi^{t_0} \middle\vert a_\pi \in \mathbb{C} \right\}\] is a six-dimensional non-commutative unital $C^\ast$-algebra. 
In general, if a von Neumann algebra $\mathcal{B} \subseteq \mathcal{A} \simeq \mathcal{M}_d(\mathbb{C})$ is a subalgebra of a finite-dimensional matrix algebra, then there exists a unitary $U$ such that
\[
\mathcal{B} = U \left( 0 \oplus \bigoplus_{k=1}^K \mathcal{M}_{d_k} \otimes \mathbbm{1}_{m_k} \right) U^\ast,
\] 
for a decomposition of the Hilbert space $\mathbb{C}^d = \mathbb{C}^{d_0} \oplus \bigoplus_{k=1}^K \mathbb{C}^{d_k} \otimes \mathbb{C}^{m_k}$, where each factor $k$ is isomorphic to a full matrix algebra of dimension $d_k^2$ which appears with multiplicity $m_k$ \cite[Chapter 3.6]{lidar_brun_2013}\nocite{bratteli_robinson_operator_algebras_1}\nocite{bratteli_robinson_operator_algebras_2}. Since von Neumann algebras and $C^\ast$-algebras coincide in finite dimensions, $\mathcal{A}^{t_0}$ is isomorphic to a sum of two one dimensional and a two dimensional matrix algebra, i.e. $6 = \sum_{k=1}^K d_k^2 = 2^2+1^2+1^2$ (Note that it cannot be a sum of six one dimensional matrix algebras, due to the non-commutativity). Using the same notation as in \cite{eggeling_werner_2001}, namely,
\begin{align*}
&X = V_{(01)}^{t_0} \ \ \text{ and } \\
&V = V_{(12)}^{t_0}=V_{(12)},
\end{align*}
with $X^\ast = X$ and $V^\ast = V$,
such that 
\begin{align*}
&\mathbbm{1}^{t_0}=\mathbbm{1}, \\
&V_{(02)}^{t_0} = VXV, \\
&V_{(012)}^{t_0} = XV, \\
&V_{(210)}^{t_0} = VX,  
\end{align*}
we get that 
\begin{align*}
&X^2=dX, \\
&V^2=\mathbbm{1} \ \ \text{ and } \\
&XVX = X.
\end{align*}
A convenient basis is then given by \cite{eggeling_werner_2001}
\begin{align*}
S_+ &= \frac{\mathbbm{1}+V}{2}\left(\mathbbm{1}- \frac{2X}{d+1}\right) \frac{\mathbbm{1}+V}{2}, \\
S_- &= \frac{\mathbbm{1}-V}{2}\left(\mathbbm{1}- \frac{2X}{d-1}\right) \frac{\mathbbm{1}-V}{2}, \\
S_0 &= \frac{1}{d^2-1}\left(d(X+VXV)-(XV+VX)\right), \\
S_1 &= \frac{1}{d^2-1} \left(d(XV+VX)-(X+VXV)\right), \\
S_2 &= \frac{1}{\sqrt{d^2-1}}\left(X-VXV\right), \\
S_3 &= \frac{i}{\sqrt{d^2-1}}\left(XV-VX\right).
\end{align*}
Denoting by $s_k(\rho):= \Tr[\rho S_k]$ for $k\in \left\{+,-,0,1,2,3\right\}$ and using the results of Eggeling and Werner \cite{eggeling_werner_2001}, we get the following criteria for complete positivity,
\[
s_+,s_-,s_0 \geq 0, \hspace{2cm} s_0^2 \geq s_1^2+s_2^2+s_3^2, \hspace{2cm}  s_+ + s_- + s_0 = 1.
\]
Translating this result back into our original notation (see Appendix~\ref{app:figures})
reduces the optimization problem given in Eq.~(\ref{eq:supremum2}) to the following convex optimization.\\
Find  
\begin{subequations}
\label{eq:optimisation3}
\begin{equation}
\sup_{\tilde{T}} \left[ z_1 d^k\left(\tilde{T}_1, \id \right) + z_2 d^k \left(\tilde{T}_2, \id \right) \right], 
\label{eq:supremum3}
\end{equation}
for $k=F, 1,2,\infty, \diamond$ with $z_1,z_2 \in \mathbb{R}$, where the supremum is taken over all quantum channels of the form
 \begin{multline}
 \tilde{T}(\rho) =a_1 d \mathbbm{1} \Tr[\rho]+a_2 d^2 \rho \otimes \frac{\mathbbm{1}}{d}+ a_3 d^2  \frac{\mathbbm{1}}{d} \otimes \rho +a_4 d \mathbb{F} \Tr[\rho] \\
 + a_5 d^2 \left( \rho \otimes \frac{\mathbbm{1}}{d}\right) \mathbb{F} + a_6 d^2 \mathbb{F} \left( \rho \otimes \frac{\mathbbm{1}}{d} \right), 
\end{multline}
with the corresponding Choi-Jamiolkowski state given by
\begin{multline}
\tau_{012}=a_1\mathbbm{1}_{012}+a_2 d \ketbra{\Omega}{\Omega}_{01}\otimes \mathbbm{1}_2 + a_3 d \ketbra{\Omega}{\Omega}_{02}\otimes \mathbbm{1}_1 \\
 +a_4 \mathbbm{1}_0 \otimes \mathbb{F}_{12} + a_5 \sum_{ijk}\ketbra{jjk}{iki} + a_6\sum_{ijk}\ketbra{kjk}{iij}, 
\end{multline}
\noindent such that
\begin{align}
0 &\leq a_1+a_4, \\
0 &\leq (a_1-a_4)\frac{1}{2}d(d-2)(d+1), \label{eq:secondIneq} \\
0 &\leq 2a_1 + (a_2+a_3)d + a_5+a_6, \\
1 &= a_1d^3+(a_2+a_3+a_4)d^2+(a_5+a_6)d, 
\end{align}
\vspace{-35pt}
\begin{multline}
-(a_2 + a_4) (a_3 + a_4) + (a_1 + a_5) (a_1 + a_6)  \\
 + (a_1 (a_2 + a_3)-   a_4 (a_5 + a_6)) d + (a_2 a_3 - a_5 a_6) d^2 \geq 0.
\label{eq:fifthIneq}
\end{multline}
\end{subequations}

In case $d>2$, it is then clear that $a_1=a_4=0$ and without loss of generality $a_5=a_6 \in \mathbb{R}$. Then the optimal cloning map and its Choi-Jamiolkowski state is given by
\begin{subequations}
\begin{align}
\tilde{T}(\rho) =&  \left( \alpha_2 \mathbbm{1} + \alpha_1 \mathbb{F}\right) \left( \rho \otimes \frac{\mathbbm{1}}{d}\right)  \left( \alpha_2 \mathbbm{1} + \alpha_1 \mathbb{F}\right), \\
\tau_{012} =&(\alpha_2)^2 \left(\ketbra{\Omega}{\Omega}_{01}\otimes \frac{\mathbbm{1}_2}{d}\right) + (\alpha_1)^2 \left(\ketbra{\Omega}{\Omega}_{02}\otimes \frac{\mathbbm{1}_1}{d} \right) \nonumber \\
 &+ \frac{\alpha_1 \alpha_2}{d^2} \sum_{ijk}\ketbra{jjk}{iki} + \frac{\alpha_1 \alpha_2}{d^2} \sum_{ijk}\ketbra{kjk}{iij}, 
\end{align}
with
\begin{equation}
(\alpha_1)^2 + (\alpha_2)^2 + \frac{2\alpha_1 \alpha_2}{d} = 1.
\end{equation}
\end{subequations}

In the case $d=2$, however, the second inequality given by Eq.~(\ref{eq:secondIneq})  vanishes. The optimization therefore does not necessarily yield the result $a_1=a_4=0$ anymore, since these might now take negative values. It turns out that this is a freedom in the parametrization, however, still yielding the same universal optimal quantum cloning channel. We may therefore state the following Theorem~\ref{thm:OptimalCloning} in full agreement with \cite{braunstein_buzek_hillery_2001, cerf_2000_QI, iblisdir_acin_cerf_filip_fiurasek_gisin_2005}, in which the optimal universal $1 \to 2$ asymmetric quantum cloning channel has been derived too. We have, however, mostly used the symmetry idea of Eggeling and Werner as well as Vollbrecht and Werner \cite{eggeling_werner_2001, vollbrecht_werner_2001} that exploit a similar symmetry property of the quantum states to study separability properties and entanglement measures. 

\noindent
\begin{theorem}[Optimal universal $1 \to 2$ asymmetric quantum cloning channel]
\label{thm:OptimalCloning}
The optimal universal $1 \to 2$ asymmetric quantum cloning channel $\tilde{T}_{\text{optimal}}: \mathcal{M}_d \to \mathcal{M}_d \otimes \mathcal{M}_d$ for any quantum state $\rho \in \mathcal{D}_d$ is given by
\begin{equation}
\tilde{T}_{\text{optimal}}(\rho)  =\left( \alpha_2 \mathbbm{1} + \alpha_1 \mathbb{F}\right) \left( \rho \otimes \frac{\mathbbm{1}}{d}\right)  \left( \alpha_2 \mathbbm{1} + \alpha_1 \mathbb{F}\right),
\label{eq:OptimalCloningChannel_thm}
\end{equation}
with $\left(\alpha_1\right)^2 + \left(\alpha_2\right)^2 + \frac{2\alpha_1 \alpha_2}{d} = 1$, $\alpha_1, \alpha_2 \in \mathbb{R}$, and where $\mathbb{F}:= \sum_{i,j=1}^d \ketbra{ji}{ij}$ is the flip (or swap) operator. \\
\end{theorem}

What is interesting to  notice is that as the dimension of the underlying Hilbert space $d$ increases, the optimal cloning map approaches the trivial approach to quantum cloning. The trivial approach is represented by the quantum channel \[T_{\text{trivial}}\left(\rho \right)=\alpha \rho \otimes \frac{\mathbbm{1}}{d} + \left(1-\alpha\right) \frac{\mathbbm{1}}{d}\otimes \rho,\] where $\alpha \in [0,1]$. Instead of cloning the quantum state $\rho$, an identity channel is applied and an additional state is prepared, the maximally mixed state. Thus, in the limit as the dimension of the underlying Hilbert space increases, $d \to \infty$, even approximate quantum cloning is not possible.

\subsection*{Determining achievable quantum clone qualities numerically}  
\noindent
In order to support our findings, it is also possible to rewrite the optimization problem given by Eq.~(\ref{eq:optimisation3}) as a semidefinite program (SDP) \cite{vandenberghe_boyd_1996}. 
To solve this semidefinite program, we used cvx, a package for specifying and solving convex programs \cite{cvx, grant_boyd_08}. 
Let
\[
z = \begin{pmatrix}
	z_1 \\ z_2
\end{pmatrix} 
\in \mathbb{R}^2 \ \text{ and } \ 
D = \begin{pmatrix}
	d^k(T_1, \id) \\ d^k(T_2, \id)
\end{pmatrix},
\]
for $k=F, 1,2,\infty, \diamond$ with $T_i\left(\cdot\right) = \alpha_i^2\frac{\mathbbm{1}}{d}\Tr \left[\cdot \right] + \left( 1-\alpha_i^2\right) \id \left( \cdot \right)$. \\ 
Maximise
\[
z^T\cdot D  
\]
subject to
\begin{multline*}
s_0 \oplus s_+ \oplus  s_- \oplus \begin{pmatrix} s_0+s_3 & s_1+is_2 \\ s_1+is_2 & s_0-s3 \end{pmatrix} \oplus  s_++s_-+s_0-1  \\
  \oplus 1-(s_++s_-+s_0) \oplus \alpha_1^2 \oplus \alpha_2^2 \oplus \frac{d^2}{d^2-1}-\alpha_1^2 \oplus \frac{d^2}{d^2-1}-\alpha_2^2 \vphantom{\begin{pmatrix} s_0+s_3 & s_1+is_2 \\ s_1+is_2 & s_0-s3 \end{pmatrix}}  \geq 0.
\end{multline*}
The corresponding analytical results for different figures of merit are shown in Figure~\ref{fig:Picture_ContourPlot_Fidelity} up to Figure~\ref{fig:Picture_ConvexHull_FrobeniusNorm} in the Appendix~\ref{app:figures}.

%\newpage
%%
%\begin{figure}[p]
	%\centering
		%\includegraphics[trim=110pt 250pt 110pt 250pt, clip, width=1.0\textwidth]{Picture_SDP_Fidelity.pdf}
	%\caption{SDP with $d^F(T_i, \id)=\bra{\Omega} \tau_{0i} \ket{\Omega}$, $i=1,2$, using {MATLAB} \cite{Matlab}.}
	%\label{fig:Picture_SDP_Fidelity}
%\end{figure}
%%
%\begin{figure}[p]
	%\centering
		%\includegraphics[trim=110pt 250pt 110pt 250pt, clip, width=1.0\textwidth]{Picture_SDP_Trace_Norm.pdf}
	%\caption{SDP with $d^1(T_i, \id)=2-\sup_\rho \norm{T_i(\rho)-\rho}_1$, $i=1,2$, using {MATLAB} \cite{Matlab}.}
	%\label{fig:Picture_SDP_Trace_Norm}
%\end{figure}
%%
%\begin{figure}[p]
	%\centering
		%\includegraphics[trim=110pt 250pt 110pt 250pt, clip, width=1.0\textwidth]{Plot_SDP_FrobeniusNorm.pdf}
	%\caption{SDP with $d^2(T_i, \id)=1-\sup_\rho \norm{T_i(\rho)-\rho}_2$, $i=1,2$, using {MATLAB} \cite{Matlab}.}
	%\label{fig:Picture_SDP_Frobenius_Norm}
%\end{figure}
%%
%\begin{figure}[p]
	%\centering
		%\includegraphics[trim=110pt 250pt 110pt 250pt, clip, width=1.0\textwidth]{Plot_SDP_OperatorNorm.pdf}
	%\caption{SDP with $d^\infty(T_i, \id)=1-\sup_\rho \norm{T_i(\rho)-\rho}_\infty$, $i=1,2$, using {MATLAB} \cite{Matlab}.}
	%\label{fig:Picture_SDP_Operator_Norm}
%\end{figure}
%%
%\newpage

%%%%%%%%%%%%%%%%%%%%%%%%%%%%%%%%%%%%%%%%%%%%%
\section{The Set of all achievable Single Quantum Clone Qualities}
\label{sec:Bipolar}
\begin{figure}[htb]
\vspace*{13pt}
	\centering
	\begin{tikzpicture}[thick]
		\draw [<->] (0,2.5) -- (0,0) -- (2.5,0);
		\draw [thin, gray, dashed] (0.5,0) -- (0.5,2);
		\draw [thin, gray, dashed] (0,0.5) -- (2,0.5);
		%\draw [thin, gray, dashed] (0,2) -- (2,2) -- (2,0);
		\draw [fill = gray, fill opacity=0.2] (0,1.5) to [out=90, in = 180] (0.5,2) to [out=0, in = 135] (1.52,1.52) to [out=-45, in = 90] (2,0.5) to [out=270, in = 0] (1.5,0) -- (0,0) -- (0,1.5);
		\draw [thin] (-1pt,2) -- (+1pt,2) node [anchor=east] {$1$};
		\draw [thin] (-1pt,1.5) -- (+1pt,1.5) node [anchor=east] {$\frac{d^2-1}{d^2}$};
		\draw [thin] (-1pt,0.5) -- (+1pt,0.5) node [anchor=east] {$\frac{1}{d^2}$};
		\draw [thin] (2,-1pt) -- (2,+1pt) node [anchor=north] {$1$};
		\draw [thin] (1.5,-1pt) -- (1.5,+1pt) node [anchor=north] {$\frac{d^2-1}{d^2}$};
		\draw [thin] (0.5,-1pt) -- (0.5,+1pt) node [anchor=north] {$\frac{1}{d^2}$};
		\node [below left] at (0,0) {$0$};
		\node [above left] at (0,2.5) {$\left\langle \Omega \right\rvert    \tau_{01}   \left\lvert \Omega \right\rangle$};
  	\node [below right] at (2.5,0) {$\left\langle \Omega \right\rvert    \tau_{02}      \left\lvert \Omega \right\rangle$};
		\node at (1,1) {$\mathcal{C}^F$};
		\end{tikzpicture}
		\vspace*{13pt}
	\caption{\label{fig:Picture_Set_Fidelity}Set of all attainable single quantum clone fidelities $\mathcal{C}^F$. The figure of merit is the single clone fidelity $d^F(T_i, \id) = \bra{\Omega} \tau_{0i} \ket{\Omega}$ for $i=1,2$, where $\tau_{0i} := \id \otimes T_i \left( \ketbra{\Omega}{\Omega} \right)$ is the Choi-Jamiolkowski state.}
\end{figure}
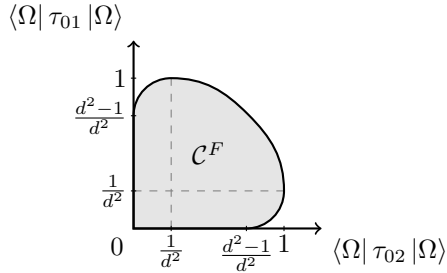

%%%%%%%%%%%%%%%%%%%%%%%%%%%%%%%%%%%%%%%%%%%%%
\noindent
In the previous part, we were only interested in the optimal asymmetric quantum cloning channel describing the boundary of the set of all achievable single quantum clone qualities. In this chapter, we analytically derive this set using different figures of merit. Let us, however, turn to the  fidelity $d^F(T_i,\id) = \bra{\Omega} \tau_{0i} \ket{\Omega}$ for $i=1,2$, where $\tau_{0i}=\id \otimes \tilde{T}_i \left( \ketbra{\Omega}{\Omega}\right)$, first, such that the optimization problem is given by
\begin{equation}
\sup_{\tilde{T}} \left[ z_1 d^F(\tilde{T}_1, \id) + z_2 d^F(\tilde{T}_2, \id) \right] = \sup_{\substack{\tau \geq 0 \\ \Tr_{12}[\tau]=\frac{\mathbbm{1}_0}{d}}} \left[ z_1 \bra{\Omega} \tau_{01} \ket{\Omega} + z_2 \bra{\Omega} \tau_{02} \ket{\Omega} \right]. 
\label{eq:OptFidelity1}
\end{equation}
This is visualized in Figure~\ref{fig:Picture_Set_Fidelity}, which shows the set of all attainable qualities of the two quantum clones,
\begin{equation}
\mathcal{C}^F = \left\{ z \in \mathbb{R}^2 \middle\vert z = \begin{pmatrix} \bra{\Omega} \tau_{01} \ket{\Omega} \\ \bra{\Omega} \tau_{02} \ket{\Omega} \end{pmatrix} \right\}.
\label{eq:CF}
\end{equation}

First of all, we notice that for $i=1,2$ and $\bar{i} = 2,1$, if the overlap of $\tau_{0i}$ with the maximally entangled state is $\bra{\Omega} \tau_{0i} \ket{\Omega} = 1$ yielding $\tau_{0i} = \ketbra{\Omega}{\Omega}_{0i}$, then the overall state must be $\tau_{012} = \ketbra{\Omega}{\Omega}_{0i} \otimes \frac{\mathbbm{1}_{\bar{i}}}{d}$, such that the other marginal state turns out to be $\tau_{0\bar{i}} = \frac{\mathbbm{1}_0}{d} \otimes  \frac{\mathbbm{1}_{\bar{i}}}{d}$. This gives $\bra{\Omega} \tau_{0\bar{i}} \ket{\Omega} = \frac{1}{d^2}$. 
Furthermore, if the overlap of $\tau_{0i}$ with the maximally entangled state is $\bra{\Omega} \tau_{0i} \ket{\Omega} =0$ yielding $\tau_{0i}= \frac{d^2}{d^2-1} \frac{\mathbbm{1}}{d^2}-\frac{1}{d^2-1}\ketbra{\Omega}{\Omega}$, then the other marginal state must be $\tau_{0\bar{i}}= \frac{1}{d^2-1} \frac{\mathbbm{1}}{d^2}+\frac{d^2-2}{d^2-1}\ketbra{\Omega}{\Omega}$, such that its overlap with the maximally entangled state is $\bra{\Omega} \tau_{0\bar{i}} \ket{\Omega} =\frac{d^2-1}{d^2}$. This gives four extreme points of our set $\mathcal{C}^F$ as illustrated in Figure~\ref{fig:Picture_Set_Fidelity}.

\noindent
\begin{lemma}
\label{lem:TP}
Let $\tau$ be the Choi-Jamiolkowski operator of a symmetrized quantum channel. Then 
\begin{equation}
\tau_{012} \geq 0 \text{ and } \Tr_{12} \left[ \tau_{012} \right] = \frac{\mathbbm{1}_0}{d} \nonumber
\end{equation}
is equivalent to
\begin{equation}
\tau_{012} \geq 0 \text{ and } \Tr \left[ \tau_{012} \right] = 1. \nonumber
\end{equation}
\end{lemma}

\noindent
\begin{proof}{
If $\Tr_{12} \left[ \tau_{012} \right] = \frac{\mathbbm{1}_0}{d}$, then taking the full trace gives $\Tr_{012} \left[ \tau_{012} \right] = 1$. The other direction follows from the form of the Choi-Jamiolkowsi state of a symmetrized quantum channel.
Using $\tau_{012} = \sum_{\pi \in \mathrm{S}_3} a_\pi V_\pi^{t_0}$ gives
\begin{align*}
&\Tr_{012} \left[ \tau_{012} \right] = 1 \\
\Leftrightarrow &\sum_{\pi \in \mathrm{S}_3} a_\pi \Tr_{012} \left[ V_\pi^{t_0} \right] = 1 \\
\Leftrightarrow & \left( a_1d^2+(a_2+a_3+a_4)d+a_5+a_6 \right)d = 1.
\end{align*}
Now
\begin{equation}
\Tr_{12} \left[ \tau_{012} \right] = \left(  a_1d^2+(a_2+a_3+a_4)d+a_5+a_6 \right) \mathbbm{1}_0 = \frac{\mathbbm{1}_0}{d}  \nonumber
\end{equation}
}
\end{proof}

In order to describe the set of all achievable single clone qualities, consider \[H_z = z_1 \ketbra{\Omega}{\Omega}_{01} \otimes \mathbbm{1}_2 + z_2 \ketbra{\Omega}{\Omega}_{02} \otimes \mathbbm{1}_1, \ \ \text{ with } z_1,z_2 \in \mathbb{R}.\] Then, one notices that our optimization problem given by Eq.~(\ref{eq:OptFidelity1}) may be rewritten using Lemma~\ref{lem:TP} as 
\begin{align*}
&\sup_{\substack{\tau \geq 0 \\ \Tr_{12}[\tau]=\frac{\mathbbm{1}_0}{d}}} \left[ z_1 \bra{\Omega} \tau_{01} \ket{\Omega} + z_2 \bra{\Omega} \tau_{02} \ket{\Omega} \right] \\
=&\sup_{\substack{\tau \geq 0 \\ \Tr[\tau]=1}} \left[ z_1 \bra{\Omega} \tau_{01} \ket{\Omega} + z_2 \bra{\Omega} \tau_{02} \ket{\Omega} \right] \\
=& \sup_{\substack{\tau \geq 0 \\ \Tr[\tau]=1}} \Tr \left[ z_1 \ketbra{\Omega}{\Omega}_{01} \otimes \mathbbm{1}_2 \tau_{012}  + z_2 \ketbra{\Omega}{\Omega}_{02} \otimes \mathbbm{1}_1 \tau_{012} \right] \\
=& \sup_{\substack{\tau \geq 0 \\ \Tr[\tau]=1}}  \Tr \left[H_z \tau_{012} \right] \\
=& \lambda_{\text{max}}\left( H_{z} \right),
\end{align*}
where $\lambda_{\text{max}}\left( H_{z} \right)$ is the maximum eigenvalue of $H_z$.
We are thus interested in the largest eigenvalue of $H_z$, denoted as $\lambda_{\text{max}}\left( H_{z} \right)$, i.e.
\[
\lambda_{\text{max}}\left(H_z\right) =  \sup_{\substack{\tau \geq 0 \\ \Tr[\tau]=1}} \left[ z_1 \bra{\Omega} \tau_{01} \ket{\Omega} + z_2 \bra{\Omega} \tau_{02} \ket{\Omega} \right] = \sup_{x\in \mathcal{C}^F} \left\langle z,x \right\rangle. 
\]
The set $\mathcal{C}^F$ defined in Eq.~(\ref{eq:CF}) may then be expressed using the notion of a polar, which is defined as follows.

\noindent
\begin{definition}[Polar {\cite[Definition 5.101]{aliprantis_border_2007}}]
\label{def:Polar}
Consider a finite dimensional vector space $X$ and its dual vector space $X^\ast$. The one-sided polar $A^\odot$ of a nonempty subset $A$ of $X$, is the subset of $X^\ast$ defined by 
\[
A^\odot := \left\{ x' \in X^\ast: \left\langle x, x' \right\rangle \leq 1 \text{ for all } x \in A \right\}.
\]
Likewise, if $B$ is a nonempty subset of $X^\ast$, then its one-sided polar is the subset of $X$ defined by 
\[
B^\odot := \left\{ x \in X: \left\langle x, x' \right\rangle \leq 1 \text{ for all } x' \in B \right\}.
\]
The one-sided bipolar of a subset $A$ of $X$ is the set $\left( A^\odot \right)^\odot$ written simply as $A^{\odot \odot}$. The bipolar of a subset of $X^\ast$ is defined in a similar manner. 
\end{definition}
Lemma~\ref{lem:PolarProperties}, which can be found in the appendix, gives some properties of the one-sided polar, in order to allow a more intuitive handling of this Definition~\ref{def:Polar}. With this definition at hand, we may state the Bipolar Theorem~\ref{thm:bipolar}.

\noindent
\begin{theorem}[Bipolar Theorem {\cite[Theorem 5.103]{aliprantis_border_2007}}]
\label{thm:bipolar}
Consider a finite dimensional vector space $X$ and its dual vector space $X^\ast$ and let $A$ be a nonempty subset of $X$. 
The one-sided bipolar $A^{\odot \odot}$ is the convex closed hull of $A \cup \{0 \}$. Hence if $A$ is convex, closed, and contains zero, then $A = A^{\odot \odot}$.
Corresponding results hold for subsets of $X^\ast$.
\end{theorem}

The Bipolar Theorem~\ref{thm:bipolar} has numerous applications in functional analysis.\footnote{Further information about the concept of a polar and a more general statement of the Bipolar theorem can be found in \cite{aliprantis_border_2007}.}  In quantum information it always presents a very powerful tool when one wishes to fully characterize a closed convex set, which is exactly what we would like to do here.
The one-sided polar $\left(\mathcal{C}^F\right)^{\odot}$ of the non-empty set $\mathcal{C}^F \subseteq \mathbb{R}^2$, defined in Eq.~(\ref{eq:CF}) is therefore given as
\begin{align*}
\left(\mathcal{C}^F\right)^{\odot} &= \left\{ z \in \mathbb{R}^2 \middle\vert \forall \ x \in \mathcal{C}^F: \left\langle z, x \right\rangle \leq 1 \right\} \\
&= \left\{ z \in \mathbb{R}^2 \middle\vert \sup_{x \in \mathcal{C}^F}\left\langle z, x \right\rangle \leq 1 \right\} \\
&=\left\{ z \in \mathbb{R}^2 \middle\vert \lambda_{\text{max}}\left(H_z\right) \leq 1 \right\}.
\end{align*}
Since the one-sided bipolar $\left(\mathcal{C}^F\right)^{\odot \odot}$ is just the one-sided polar of the one-sided polar, we get
\begin{align*}
\left(\mathcal{C}^F\right)^{\odot \odot} &=  \left\{ x \in \mathbb{R}^2 \middle\vert \forall \ z \in \left(\mathcal{C}^F\right)^{\odot}: \left\langle x,z \right\rangle \leq 1 \right\} \\
&=  \left\{ x \in \mathbb{R}^2 \middle\vert \forall \ z \in \mathbb{R}^2: \text{ if } \lambda_{\text{max}}\left(H_z\right) \leq 1 \text{ then }  \left\langle x,z \right\rangle \leq 1 \right\} \\
&= \left\{ x \in \mathbb{R}^2 \middle\vert \forall \ z \in \mathbb{R}^2: \left\langle x,\frac{z}{\lambda_{\text{max}}\left(H_z\right)} \right\rangle \leq 1 \right\} \\
&= \left\{ x \in \mathbb{R}^2 \middle\vert \sup_{z\in \mathbb{R}^2} \left\langle x,\frac{z}{\lambda_{\text{max}}\left(H_z\right)} \right\rangle \leq 1 \right\}.
\end{align*}
%
%\vspace{15pt}
%\begin{figure}[htp]
	%\centering
		%\includegraphics[width=\textwidth]{Picture_Halfplanes.pdf}
	%\caption{Differentiating the three cases.}
	%\label{fig:Picture_Halfplanes}
%\end{figure}
In order to analyze this even further, one may now realize that every vector $\left(\begin{smallmatrix} z_1 \\ z_2 \end{smallmatrix}\right) \in \mathbb{R}^2$ can be written as
$b \left(\begin{smallmatrix} v \\ 1-v \end{smallmatrix}\right)$, with $b \in \mathbb{R}_+$ and $v \in \mathbb{R}$, if $z_1+z_2 >0$, or $b \left(\begin{smallmatrix} -v \\ v-1 \end{smallmatrix}\right)$, with $b \in \mathbb{R}_+$  and $v \in \mathbb{R}$, if $z_1+z_2 <0$, or $\left(\begin{smallmatrix} \pm v \\ \mp v \end{smallmatrix}\right)$, with $v \in \mathbb{R}$, if $z_1+z_2 =0$.
% as illustrated in Figure \ref{fig:Picture_Halfplanes}. 
This is helpful, because $b\in \mathbb{R}_+$ and $\lambda_{\text{max}}\left(H_{bz}\right) = b\lambda_{\text{max}}\left(H_z\right)$.
Now differentiating these three cases, %as shown in Figure \ref{fig:Picture_Halfplanes}
the one-sided bipolar is 
\begin{align*}
\left(\mathcal{C}^F\right)^{\odot \odot} = \left\{ \vphantom{\frac{\left\langle x,\begin{pmatrix} v \\ 1-v \end{pmatrix} \right\rangle}{\lambda_{\text{max}}\left(H_{\left(\begin{smallmatrix} v \\ 1-v \end{smallmatrix}\right)}\right)}} x \in \mathbb{R}^2 \right\vert & \sup_{v\in \mathbb{R}^2} \frac{\left\langle x,\begin{pmatrix} v \\ 1-v \end{pmatrix} \right\rangle}{\lambda_{\text{max}}\left(H_{\left(\begin{smallmatrix} v \\ 1-v \end{smallmatrix}\right)}\right)}  \leq 1  \\
\wedge & \sup_{v\in \mathbb{R}^2} \frac{\left\langle x,\begin{pmatrix} -v \\ v-1 \end{pmatrix}  \right\rangle}{\lambda_{\text{max}}\left(H_{\left(\begin{smallmatrix} -v \\ v-1 \end{smallmatrix}\right)}\right)} \leq 1  \\
\wedge & \sup_{v\in \mathbb{R}^2} \frac{\left\langle x,\begin{pmatrix} \pm v \\ \mp v \end{pmatrix} \right\rangle}{\lambda_{\text{max}}\left(H_{\left(\begin{smallmatrix} \pm v \\  \mp v \end{smallmatrix}\right)}\right)}  \leq 1  \left. \vphantom{\frac{\left\langle x,\begin{pmatrix} v \\ 1-v \end{pmatrix} \right\rangle}{\lambda_{\text{max}}\left(H_{\left(\begin{smallmatrix} \pm v \\ \mp v \end{smallmatrix}\right)}\right)}} \right\}. 
\end{align*}
Note that by analyzing the rank of $H_z$ we always expect an eigenvalue equal to zero. Therefore, the maximum eigenvalue must always be non-negative, i.e. $\lambda_{\text{max}}\left(H_{z}\right) \geq 0$. It turns out that
\begin{align*}
\lambda_{\text{max}}\left(H_{\left(\begin{smallmatrix} v \\ 1-v \end{smallmatrix}\right)}\right) &= \frac{1}{2d}\left(d+\sqrt{d^2+4(d^2-1)(v-1)v} \right), \\
\lambda_{\text{max}}\left(H_{\left(\begin{smallmatrix} -v \\ v-1 \end{smallmatrix}\right)}\right) &= 
\begin{cases}
0 & \text{if } 0 \geq v \geq 1, \\
\frac{1}{2d}\left(-d+\sqrt{d^2+4(d^2-1)(v-1)v} \right) & \text{otherwise},
\end{cases} \\
\lambda_{\text{max}}\left(H_{\left(\begin{smallmatrix} \pm v \\ \mp v \end{smallmatrix}\right)}\right) &= v\sqrt{\frac{d^2-1}{d^2}}.
\end{align*}

\noindent
\begin{proposition}[Convexity]
\label{prop:CFconvex}
The set
\[
\mathcal{C}^F = \left\{ z \in \mathbb{R}^2 \middle\vert z = \begin{pmatrix} \bra{\Omega} \tau_{01} \ket{\Omega} \\ \bra{\Omega} \tau_{02} \ket{\Omega} \end{pmatrix} \right\}
\]
is convex.
\end{proposition}

\noindent
\begin{proof}{
Let $z^A, z^B \in \mathcal{C}$, then for $\lambda \in [0,1]$, $z^C = \lambda z^A + (1-\lambda)z^B \in \mathcal{C}^F$, because
\begin{align*}
z^C &= \lambda z^A + (1-\lambda)z^B \\
&= \lambda \begin{pmatrix} \bra{\Omega} \tau^A_{01} \ket{\Omega} \\ \bra{\Omega} \tau^A_{02} \ket{\Omega} \end{pmatrix} + (1-\lambda) \begin{pmatrix} \bra{\Omega} \tau^B_{01} \ket{\Omega} \\ \bra{\Omega} \tau^B_{02} \ket{\Omega} \end{pmatrix} \\
%&= \begin{pmatrix} \bra{\Omega} \lambda \tau^A_{01} + (1-\lambda) \tau^B_{01} \ket{\Omega} \\ \bra{\Omega} \lambda \tau^A_{02} + (1-\lambda) \tau^B_{02} \ket{\Omega}\end{pmatrix} \\
%&=\begin{pmatrix} \bra{\Omega} \lambda \id \otimes \tilde{T}^A_{1}+ (1-\lambda) \id \otimes \tilde{T}^B_{1}  \left( \ketbra{\Omega}{\Omega} \right) \ket{\Omega} \\ \bra{\Omega} \lambda \id \otimes \tilde{T}^A_{2}+ (1-\lambda) \id \otimes \tilde{T}^B_{2}  \left( \ketbra{\Omega}{\Omega} \right) \ket{\Omega}\end{pmatrix} \\
%&=\begin{pmatrix} \bra{\Omega} \id \otimes \left( \lambda \tilde{T}^A_{1}+ (1-\lambda)  \tilde{T}^B_{1} \right)  \left( \ketbra{\Omega}{\Omega} \right) \ket{\Omega} \\ \bra{\Omega}\id \otimes \left( \lambda \tilde{T}^A_{2}+ (1-\lambda) \tilde{T}^B_{2} \right)  \left( \ketbra{\Omega}{\Omega} \right)\ket{\Omega}\end{pmatrix} \\
& = \begin{pmatrix} \bra{\Omega} \tau^C_{01} \ket{\Omega} \\ \bra{\Omega} \tau^C_{02} \ket{\Omega} \end{pmatrix} \in \mathcal{C}^F
\end{align*}}
\end{proof}
By using the Bipolar Theorem~\ref{thm:bipolar} together with the fact that $\mathcal{C}^F$ is convex, as shown in Proposition~\ref{prop:CFconvex}, closed and contains the origin, we see that
$\mathcal{C}^F = \left(\mathcal{C}^F\right)^{\odot \odot}$.
A cumbersome computation then shows that the boundary of this set is described by 
\begin{multline}
\frac{1}{d+1} \left( \sqrt{\bra{\Omega} \tau_{01} \ket{\Omega}} + \sqrt{\bra{\Omega} \tau_{02} \ket{\Omega}}\right)^2    \\
+  \frac{1}{d-1}\left( \sqrt{\bra{\Omega} \tau_{01} \ket{\Omega}} - \sqrt{\bra{\Omega} \tau_{02} \ket{\Omega}}\right)^2 = \frac{2}{d},
\label{eq:UpBoundary}
\end{multline}

which is illustrated in Figure~\ref{fig:Picture_ContourPlot_Fidelity}. We may therefore state the following Theorem~\ref{thm:OptimalCloningSet}, summarizing the main result. 

\noindent
\begin{theorem}[Set of all attainable single clone fidelities within universal $1 \to 2$ asymmetric quantum cloning]
\label{thm:OptimalCloningSet}
The set of all attainable clone qualities in terms of single clone fidelities $d^F\left(T_i, \id \right) = \bra{\Omega}\tau_{0i}\ket{\Omega}$ with $i=1,2$, where $\tau_{0i}:= \id \otimes \tilde{T}_i \left(\ketbra{\Omega}{\Omega}\right)$ is the Choi-Jamiolkowski state of the marginals of the optimal quantum cloning channel, given by Eq.~(\ref{eq:OptimalCloningChannel_thm}), with 
$\ketbra{\Omega}{\Omega} := \frac{1}{d} \sum_{i,j=1}^{d} \ketbra{ii}{jj}$ being the maximally entangled state,
 is given by 
\begin{subequations}
\begin{align}
\mathcal{C}^F = \left\{ \vphantom{\frac{\left\langle x,\begin{pmatrix} v \\ 1-v \end{pmatrix} \right\rangle}{\lambda_{\text{max}}\left(H_{\left(\begin{smallmatrix} v \\ 1-v \end{smallmatrix}\right)}\right)}} x \in \mathbb{R}^2 \right\vert & \sup_{v\in \mathbb{R}^2} \frac{\left\langle x,\begin{pmatrix} v \\ 1-v \end{pmatrix} \right\rangle}{\lambda_{\text{max}}\left(H_{\left(\begin{smallmatrix} v \\ 1-v \end{smallmatrix}\right)}\right)}  \leq 1 \nonumber  \\
\wedge & \sup_{v\in \mathbb{R}^2} \frac{\left\langle x,\begin{pmatrix} -v \\ v-1 \end{pmatrix}  \right\rangle}{\lambda_{\text{max}}\left(H_{\left(\begin{smallmatrix} -v \\ v-1 \end{smallmatrix}\right)}\right)} \leq 1  \nonumber \\
\wedge & \sup_{v\in \mathbb{R}^2} \frac{\left\langle x,\begin{pmatrix} \pm v \\ \mp v \end{pmatrix} \right\rangle}{\lambda_{\text{max}}\left(H_{\left(\begin{smallmatrix} \pm v \\  \mp v \end{smallmatrix}\right)}\right)}  \leq 1  \left. \vphantom{\frac{\left\langle x,\begin{pmatrix} v \\ 1-v \end{pmatrix} \right\rangle}{\lambda_{\text{max}}\left(H_{\left(\begin{smallmatrix} \pm v \\ \mp v \end{smallmatrix}\right)}\right)}} \right\}, 
\label{eq:C_thm}
\end{align}
where $\lambda_{\text{max}}\left(H_z\right)$ denotes the maximum eigenvalue of 
\[
H_z = z_1 \ketbra{\Omega}{\Omega}_{01} \otimes \mathbbm{1}_2 + z_2 \ketbra{\Omega}{\Omega}_{02} \otimes \mathbbm{1}_1,
\]
given by
\begin{align*}
\lambda_{\text{max}}\left(H_{\left(\begin{smallmatrix} v \\ 1-v \end{smallmatrix}\right)}\right) &= \frac{1}{2d}\left(d+\sqrt{d^2+4(d^2-1)(v-1)v} \right), \\
\lambda_{\text{max}}\left(H_{\left(\begin{smallmatrix} -v \\ v-1 \end{smallmatrix}\right)}\right) &= 
\begin{cases}
0 & \text{ if } 0 \geq v \geq 1, \\
\frac{1}{2d}\left(-d+\sqrt{d^2+4(d^2-1)(v-1)v} \right) & \text{ otherwise},
\end{cases} \\
\lambda_{\text{max}}\left(H_{\left(\begin{smallmatrix} \pm v \\ \mp v \end{smallmatrix}\right)}\right) &= v\sqrt{\frac{d^2-1}{d^2}}.
\end{align*}
The upper boundary of this set is described by
\begin{multline}\label{eq:UpperBoundary_thm}
\frac{1}{d+1} \left( \sqrt{\bra{\Omega} \tau_{01} \ket{\Omega}} + \sqrt{\bra{\Omega} \tau_{02} \ket{\Omega}}\right)^2  \\
+  \frac{1}{d-1}\left( \sqrt{\bra{\Omega} \tau_{01} \ket{\Omega}} - \sqrt{\bra{\Omega} \tau_{02} \ket{\Omega}}\right)^2 = \frac{2}{d},
\end{multline}
\end{subequations}
and illustrated in Figure~\ref{fig:Picture_ContourPlot_Fidelity}, which can be found in the appendix.
\end{theorem}

This theorem is in agreement with previously established results \cite{jiang_yu_2012, studzinski_cwiklinski_horodecki_mozrzymas_2014}. Here, the authors have used a group theoretic approach, whereas our main technique comes from convex analysis. 

A similar theorem may be stated for different figures of merit, since the set of attainable  single clone qualities is convex for any $d(T_i, \id)$, satisfying the properties discussed earlier, as shown in the following proposition.

\noindent
\begin{proposition}[Convexity]
The set 
\[
\mathcal{C} = \left\{ z \in \mathbb{R}^2 \middle\vert z= \begin{pmatrix} d\left( T_1, \id\right) \\ d\left(T_2, \id \right) \end{pmatrix} \right\}
\]
is convex.
\end{proposition} 

\noindent
\begin{proof}{
Let $z^A, z^B \in \mathcal{C}$, then for $\lambda \in [0,1]$, we get
\begin{align*}
& \lambda \begin{pmatrix} d\left(T_1^A, \id \right) \\ d\left(T_2^A, \id \right) \end{pmatrix} + \left( 1 - \lambda \right) \begin{pmatrix} d\left(T_1^B, \id \right) \\ d\left(T_2^B, \id \right) \end{pmatrix} \\
= & \begin{pmatrix}
\lambda  d\left(T_1^A, \id \right) + \left( 1 - \lambda \right) d\left(T_1^B, \id \right) \\
\lambda  d\left(T_2^A, \id \right) + \left( 1 - \lambda \right) d\left(T_2^B, \id \right)
\end{pmatrix} \\
\leq & \begin{pmatrix}
	d\left(\lambda T_1^A  + \left( 1 - \lambda \right) T_1^B, \id \right) \\
	d\left(\lambda T_2^A  + \left( 1 - \lambda \right) T_2^B, \id \right)
\end{pmatrix} \\
= & \begin{pmatrix}
	d\left(T_1^C, \id \right) \\
	d\left(T_2^C, \id \right)
\end{pmatrix} \in \mathcal{C}.
\end{align*}
In the case, where the figure of merit is given by the fidelity, we even get equality. In all other cases for the lower boundary consider the following quantum channel,
\begin{align*}
T\left( \rho \right) &= \ketbra{\psi}{\psi} \otimes \left( \lambda \rho + \left( 1- \lambda \right) \ketbra{\psi}{\psi} \right), \\
\intertext{such that} 
T_i \left( \rho \right) &= \ketbra{\psi}{\psi} \text{ and } \\
T_{\bar{i}} (\rho) &= \lambda \rho + (1-\lambda) \ketbra{\psi}{\psi},
\end{align*}
for $i=1,2$ with $\lambda \in [0,1]$ and $\ket{\psi} \in \mathcal{D}_d$ some pure quantum state}
\end{proof}

This immediately gives rise to the following corollary.

\noindent
\begin{corollary}[Set of all attainable single quantum clone qualities within universal $1 \to 2$ asymmetric quantum cloning using different figures of merit]
\label{cor:QuantumCloningSet_all}
The set of all attainable single quantum clone qualities in terms of the different figures of merit $d^k(T_i, \id)$ with $i=1,2$ for $k=F,1,2,\infty, \diamond$, 
is given by 
\begin{subequations}
\begin{equation}
\mathcal{C}^k = \text{conv}\left( \left\{ 0 \right\} \cup \left\{ x^{(k)}_\text{max}  \right\} \cup  \left\{ x^{(k)}  \middle\vert g\left( f^k\left(x_1^{(k)}\right),f^k\left(x_2^{(k)}\right)\right)=0 \right\}    \right),
\label{eq:Ck}
\end{equation}
where $\left\{ 0 \right\}$ is the origin, $\left\{ x_\text{max}^{(k)} \right\}$ are the two points where $d^k(T_i, \id)$ reaches its maximum for $i=1,2$ and where the function $g:\mathbb{R}^2 \to \mathbb{R}$ is
\begin{equation}
g(x_1,x_2)= \frac{1}{d+1} \left( \sqrt{x_1} + \sqrt{x_2} \right)^2 +\frac{1}{d-1} \left( \sqrt{x_1} - \sqrt{x_2} \right)^2 - \frac{2}{d},
\label{eq:UpperBoundary_Function_k}
\end{equation}
with the functions $f^k:\mathbb{R} \to \mathbb{R}$ specified by
\begin{align}
f^F\left(x_i^{(F)}\right) &= x_i^{(F)}, \\
f^1\left(x_i^{(1)}\right) &= 1+ \frac{1+d}{d}\left(x_i^{(1)}-1\right), \\
f^2\left(x_i^{(2)}\right) &= 1+ \frac{d^2-1}{d^2}\sqrt{\frac{d}{d-1}}\left(x_i^{(2)}-1\right), \\
f^\infty \left(x_i^{(\infty)}\right) &= 1+ \frac{1+d}{d}\left(x_i^{(\infty)}-1\right) \text{ and } \\
f^\diamond \left(x_i^{(\diamond)}\right) &= x_i^{(\diamond)}. 
\end{align}
\end{subequations}
The sets are depicted in Figure~
\ref{fig:Picture_ContourPlot_Fidelity} up to Figure~\ref{fig:Picture_ConvexHull_FrobeniusNorm}
in the appendix.
\end{corollary}

%%%%%%%%%%%%%%%%%%%%%%%%%%%%
\section{Summary}
%%%%%%%%%%%%%%%%%%%%%%%%%%%
\noindent
This paper revisits the universal asymmetric $1 \to 2$ quantum cloning problem.  We derived the optimal universal $1 \to 2$ asymmetric quantum cloning channel using its symmetry properties in Theorem~\ref{thm:OptimalCloning}. Additionally, we noticed that its inherent optimization problem can be recast as a semidefinite program. This result has been derived previously by  \cite{braunstein_buzek_hillery_2001, cerf_2000_QI}.

Furthermore, we completely characterize the set of all attainable single quantum clone qualities within universal asymmetric $1 \to 2$ quantum cloning for different figures of merit in Theorem~\ref{thm:OptimalCloningSet}  and Corollary~\ref{cor:QuantumCloningSet_all} using the concept of a one-sided polar together with the famous Bipolar Theorem~\ref{thm:bipolar} from convex analysis. This is an alternative approach to the one chosen in \cite{studzinski_cwiklinski_horodecki_mozrzymas_2014}, where the authors use a general group representation approach and only study the fidelity. 

\section*{Acknowledgements}
\noindent
I would like to thank Michael M. Wolf for his constant support and insightful discussions as well as Ren\'{e} Schwonnek and David Reeb for all their help. Furthermore, I would like to thank Sabina Alazzawi and Daniel Stilck Fran{\c c}a  for all the useful comments.\\
This work is supported by the Elite Network of Bavaria through the PhD programme of excellence \textit{Exploring Quantum Matter}. 

\nocite{iblisdir_acin_cerf_filip_fiurasek_gisin_2005, kay_ramanathan_kaszlikowski_2013, studzinski_cwiklinski_horodecki_mozrzymas_2014, fiurasek_filip_cerf_2005, scarani_iblisdir_gisin_2005, Keyl_Fundamentals_2002, Keyl_Habil_2003, buzek_hillery_1996, keyl_werner_1999, cwiklinski_horodecki_studzinski_2012, buzek_hillery_bednik_1998, cerf_fiurasek_2006, cerf_2000_QI, rockafellar_1996, heinosaari_ziman_2012, Buzek_Hillery_1998}

\addcontentsline{toc}{section}{Bibliography} 
\bibliographystyle{amsalpha} %phjcp  %apalike
\bibliography{Universal_Asymmetric_Quantum_Cloning}

\appendix

\section{Proofs}
\label{appendix:proofs}

\noindent
\subsection{Proof of Lemma~\ref{lem:equiv}}
\label{app:lemmaequiv}
\noindent
\TwirlComm*

\noindent
\begin{proof}{
If
\[
\left[ \tau, \bar{U} \otimes U \otimes U \right] = 0 
\]
then
\begin{align*}
  &\int_{\mathcal{U}\left(d\right)} \left( \bar{U} \otimes U \otimes U \right) \tau \left( \bar{U} \otimes U \otimes U \right)^\ast \,\mathrm{d} U \\
	= &\int_{\mathcal{U}\left(d\right)}  \tau  \left( \bar{U} \otimes U \otimes U \right) \left( \bar{U} \otimes U \otimes U \right)^\ast  \,\mathrm{d} U \\
	= &\int_{\mathcal{U}\left(d\right)} \tau  \,\mathrm{d} U \\
	= &\tau.
\end{align*}
The other direction follows from 
\[
\int_{\mathcal{U}\left(d\right)} \left( \bar{U} \otimes U \otimes U \right) \tau \left( \bar{U} \otimes U \otimes U \right)^\ast \,\mathrm{d} U = \tau,
\]
because then for unitary $V \in \mathcal{U}\left(d\right)$ we have
\begin{align*}
&\tau \left( \bar{V} \otimes V \otimes V \right) \\
= &\int_{\mathcal{U}\left(d\right)} \left( \bar{U} \otimes U \otimes U \right) \tau \left( \bar{U} \otimes U \otimes U \right)^\ast \,\mathrm{d} U \left( \bar{V} \otimes V \otimes V \right) \\
= & \left( \bar{V} \otimes V \otimes V \right) \left( \bar{V} \otimes V \otimes V \right)^\ast  \\
& \phantom{{}=1} \int_{\mathcal{U}\left(d\right)} \left( \bar{U} \otimes U \otimes U \right) \tau \left( \bar{U} \otimes U \otimes U \right)^\ast \,\mathrm{d} U \left( \bar{V} \otimes V \otimes V \right) \\
= & \left( \bar{V} \otimes V \otimes V \right) \int_{\mathcal{U}\left(d\right)} \left( \bar{U} \otimes U \otimes U \right) \tau \left( \bar{U} \otimes U \otimes U \right)^\ast \,\mathrm{d}  U \\
= & \left( \bar{V} \otimes V \otimes V \right) \tau,
\end{align*}
where we have used the invariance property}
\end{proof}

\subsection{Proof of Proposition~\ref{prop:UUU}}
\label{app:propUUU}
\noindent
\UUU*

\noindent
\begin{proof}{
\begin{align*}
& \int_{\mathcal{U}\left(d\right)} \left( U \otimes U \otimes U \right) \tau^{t_0} \left( U \otimes U \otimes U \right)^\ast \,\mathrm{d}U \\
= & \int_{\mathcal{U}\left(d\right)} \left( U \otimes U \otimes U \right) \left(  \id \otimes \tilde{T}   \ketbra{\Omega}{\Omega}  \right)^{t_0} \left( U \otimes U \otimes U \right)^\ast \,\mathrm{d}U \\
= & \int_{\mathcal{U}\left(d\right)} \left[ \left( \left( U^\ast \right)^T \otimes U \otimes U \right)  \id \otimes \tilde{T}   \ketbra{\Omega}{\Omega} \left( U^T \otimes U^\ast \otimes U^\ast \right) \right]^{t_0} \,\mathrm{d}U \\
= & \int_{\mathcal{U}\left(d\right)} \left[ \left(  \bar{U} \otimes U \otimes U \right) \id \otimes \tilde{T}  \ketbra{\Omega}{\Omega} \left( \bar{U} \otimes U \otimes U \right)^\ast \right]^{t_0} \,\mathrm{d}U \\
\intertext{remembering that $U \otimes \bar{U} \ket{\Omega} = \ket{\Omega}$ gives} 
= &  \int_{\mathcal{U}\left(d\right)} \left[ \left(  \mathbbm{1} \otimes U \otimes U \right)  \id \otimes \tilde{T}  \left( \mathbbm{1} \otimes U^\ast \right) \ketbra{\Omega}{\Omega}  \left( \mathbbm{1} \otimes U \right)  \left( \mathbbm{1} \otimes U^\ast \otimes U^\ast \right) \right]^{t_0} \,\mathrm{d}U \\ 
= & \int_{\mathcal{U}\left(d\right)} \tau^{t_0} \, \mathrm{d}U \\
= & \tau^{t_0}.
\end{align*}
Application of Lemma~\ref{lem:equiv} finishes the proof}
\end{proof}

\section{Appendix and Figures}
\label{app:figures}
\noindent
Relation between different notation used in Chapter~\ref{sec:OptimalMap}.
%\begin{landscape}
%\[
%\begin{matrix}
%& a_1 & a_2 & a_3 & a_4 & a_5 & a_6 \\
%s_0= & a_1 2d & +a_2d^2 & + a_3d^2 & & +a_5d & + a_6 d \\
%s_1= & &a_2 d & +a_3 d & + a_4 2d & + a_5 d^2 & + a_6 d^2 \\
%s_2 = & & a_2 d \sqrt{d^2-1} & +a_3 (-d) \sqrt{d^2-1} & & & \\
%s_3 = & & & & & a_5 (-id) \sqrt{d^2-1} & + a_6 id \sqrt{d^2-1} \\ 
%s_+= & a_1 \frac{1}{2}d(d+2)(d-1) & & &+a_4 \frac{1}{2}d(d+2)(d-1) & & \\
%s_-= & a_1 \frac{1}{2}d(d-2)(d+1) & & &+a_4 \frac{1}{2}d(-d+2)(d+1) & &
%\end{matrix}
%\]
%\end{landscape}
\begin{align*}
s_0 &=  a_1 2d  +a_2d^2  + a_3d^2  +a_5d  + a_6 d \\
s_1 &= a_2 d  +a_3 d  + a_4 2d  + a_5 d^2  + a_6 d^2 \\
s_2 &=  a_2 d \sqrt{d^2-1}  +a_3 (-d) \sqrt{d^2-1}  \\
s_3 &= a_5 (-id) \sqrt{d^2-1}  + a_6 id \sqrt{d^2-1} \\ 
s_+ &=  a_1 \frac{1}{2}d(d+2)(d-1) +a_4 \frac{1}{2}d(d+2)(d-1)  \\
s_- &=  a_1 \frac{1}{2}d(d-2)(d+1) +a_4 \frac{1}{2}d(-d+2)(d+1) 
\end{align*}

\noindent
\begin{lemma}[Properties of polars {\cite[Lemma 5.102]{aliprantis_border_2007}}]
\label{lem:PolarProperties}
Consider a finite dimensional vector space $X$ and its dual vector space $X^\ast$. Let $A,B$ be nonempty subsets of $X$ and let $\left\{ A_i \right\}$ be a family of nonempty subsets of $X$. Then the following properties hold:
\begin{enumerate}
	\item If $A \subset B$, then $A^{\odot} \supset A^{\odot}$.
	\item If $\varepsilon \neq 0$, then $\left( \varepsilon A \right)^{\odot} = \frac{1}{\varepsilon} A^{\odot}$.
	\item $\cap \left( A_i^{\odot} \right) = \left( \cup A_i \right)^{\odot}$.
	\item The one-sided polar $A^{\odot}$ is nonempty, convex, closed and contains the origin. 
\end{enumerate}
The corresponding dual statements are true for subsets of $X^\ast$.
\end{lemma}

\begin{figure}[p]
\vspace*{13pt}
	\centering
		\includegraphics[trim=110pt 250pt 110pt 250pt, clip, width=1\textwidth]{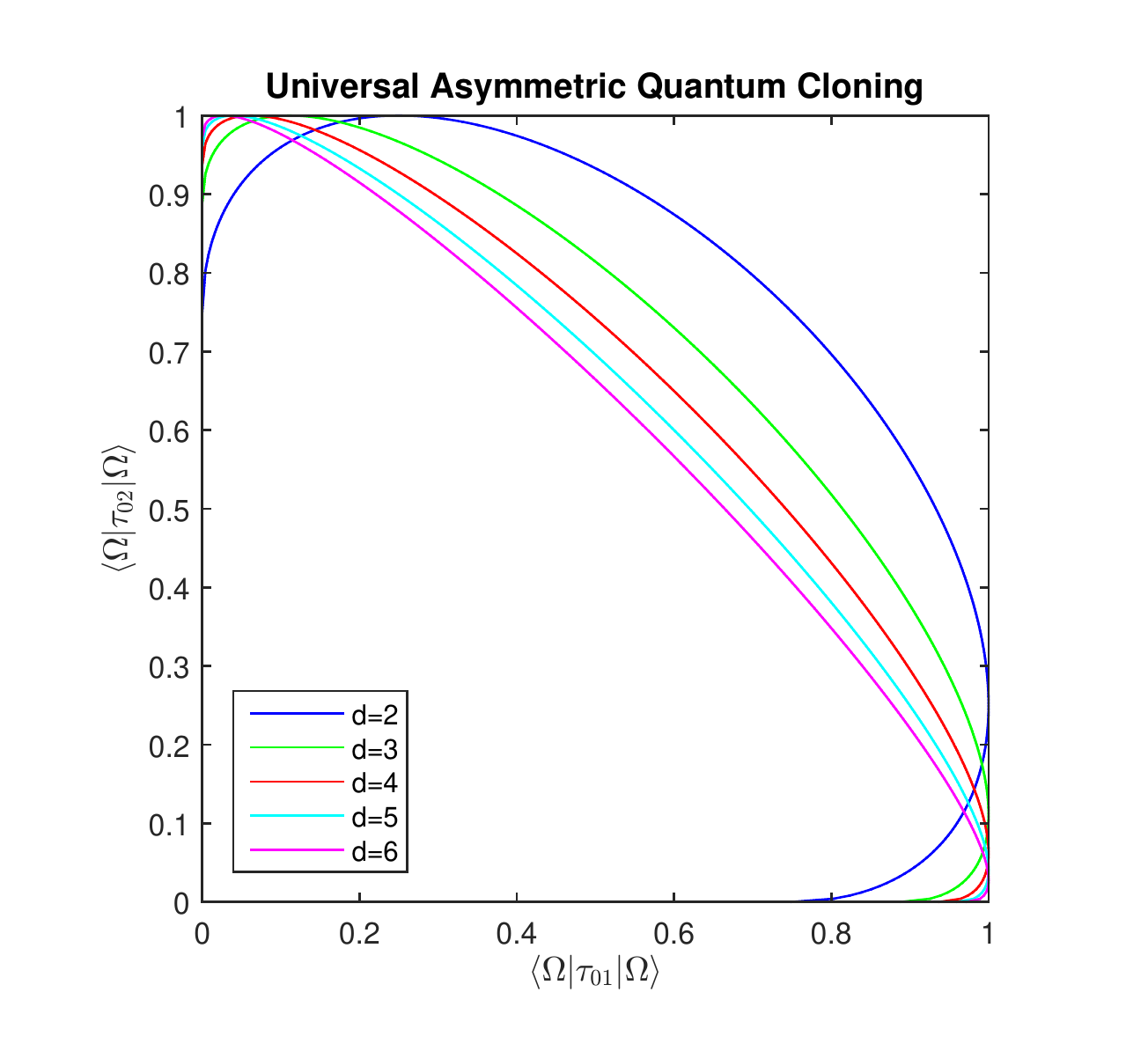}
		\vspace*{13pt}
	\caption[The set of all attainable single quantum clone qualities in terms of the fidelity $d^F(T_i, \id)=\bra{\Omega} \tau_{0i} \ket{\Omega}$, $i=1,2$, given by Eq. (\ref{eq:C_thm}) for different dimensions of the underlying Hilbert space, using {MATLAB} \cite{Matlab}.]{\label{fig:Picture_ContourPlot_Fidelity}The set of all attainable single quantum clone qualities in terms of $d^F(T_i, \id)=\bra{\Omega} \tau_{0i} \ket{\Omega}$, $i=1,2$, given by Eq.~(\ref{eq:C_thm}) for different dimensions of the underlying Hilbert space, using {MATLAB} \cite{Matlab}. The upper boundary of this set is given by 
	\begin{minipage}{\linewidth}
\begin{equation*}
\frac{1}{d+1} \left( \sqrt{x_1^{(F)}} + \sqrt{x_2^{(F)}} \right)^2   +\frac{1}{d-1} \left( \sqrt{x_1^{(F)}} - \sqrt{x_2^{(F)}} \right)^2 =\frac{2}{d}.
\end{equation*}
\end{minipage}
}

\end{figure}

	\centering
\begin{figure}[p]
\vspace*{13pt}
\centering
		\includegraphics[trim=110pt 250pt 110pt 250pt, clip, width=1\textwidth]{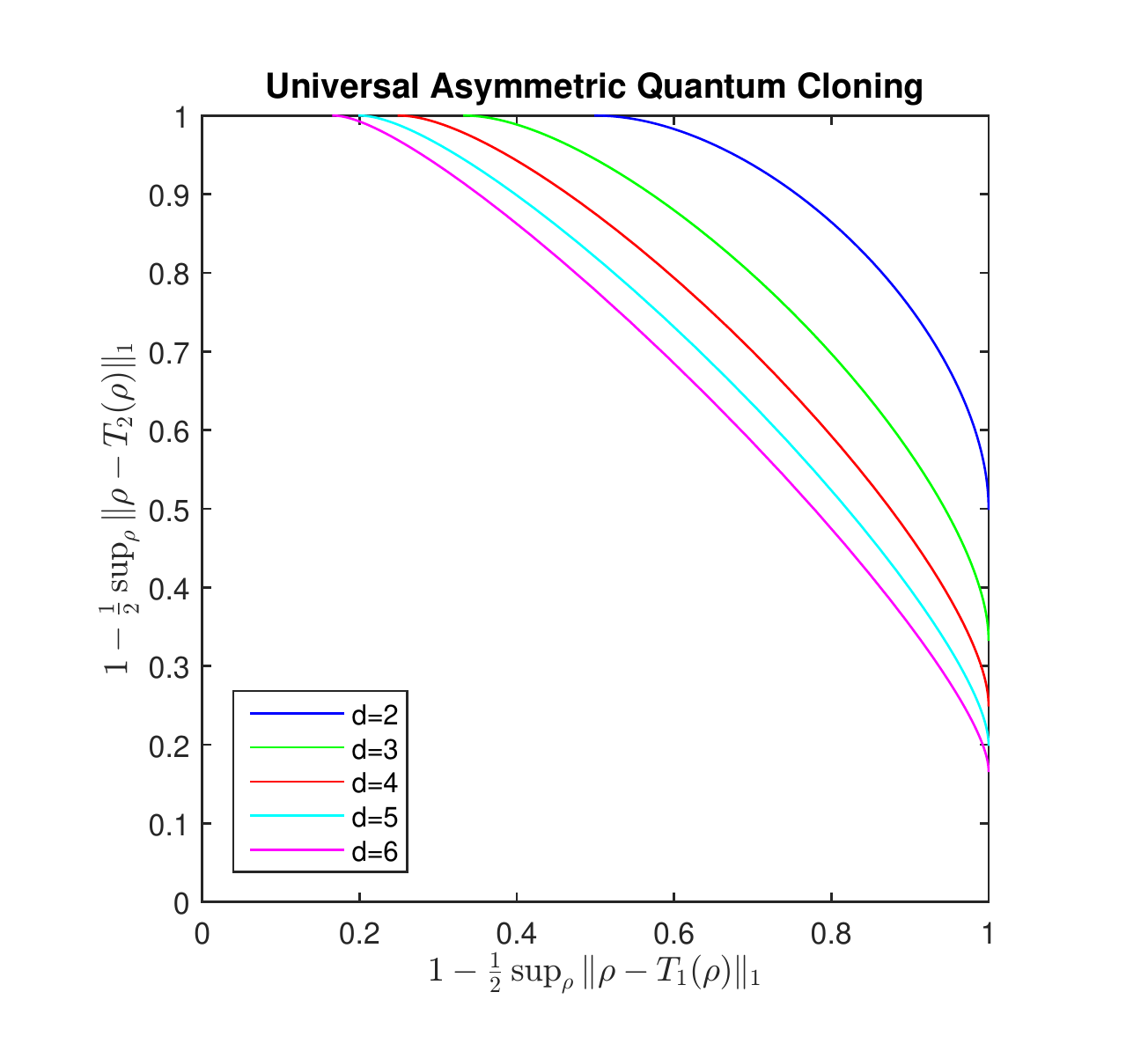}
		\vspace*{13pt}
	\caption[The set of all attainable single quantum clone qualities in terms of $d^1(T_i, \id)=1-\frac{1}{2}\sup_\rho \norm{T_i(\rho)-\rho}_1$, $i=1,2$, given by Eq. (\ref{eq:Ck}) for different dimensions of the underlying Hilbert space, using {MATLAB} \cite{Matlab}.]{\label{fig:Picture_ConvexHull_TraceNorm} The set of all attainable single quantum clone qualities in terms of $d^1(T_i, \id)=1-\frac{1}{2}\sup_\rho \norm{T_i(\rho)-\rho}_1$, $i=1,2$, given by Eq.~(\ref{eq:Ck}) for different dimensions of the underlying Hilbert space, using {MATLAB} \cite{Matlab}. The upper boundary of this set is given by
\begin{minipage}{\linewidth}
\begin{multline*}
\frac{1}{d+1} \left( \sqrt{1+\frac{1+d}{d}\left(x^{(1)}_1-1\right)} + \sqrt{1+\frac{1+d}{d}\left(x^{(1)}_2-1\right)} \right)^2  \\ +\frac{1}{d-1} \left( \sqrt{1+\frac{1+d}{d}\left(x^{(1)}_1-1\right)} - \sqrt{1+\frac{1+d}{d}\left(x^{(1)}_2-1\right)} \right)^2 =\frac{2}{d}.
\end{multline*}
\end{minipage}}
\end{figure}

\begin{figure}[p]
\vspace*{13pt}
	\centering
		\includegraphics[trim=110pt 250pt 110pt 250pt, clip, width=1.0\textwidth]{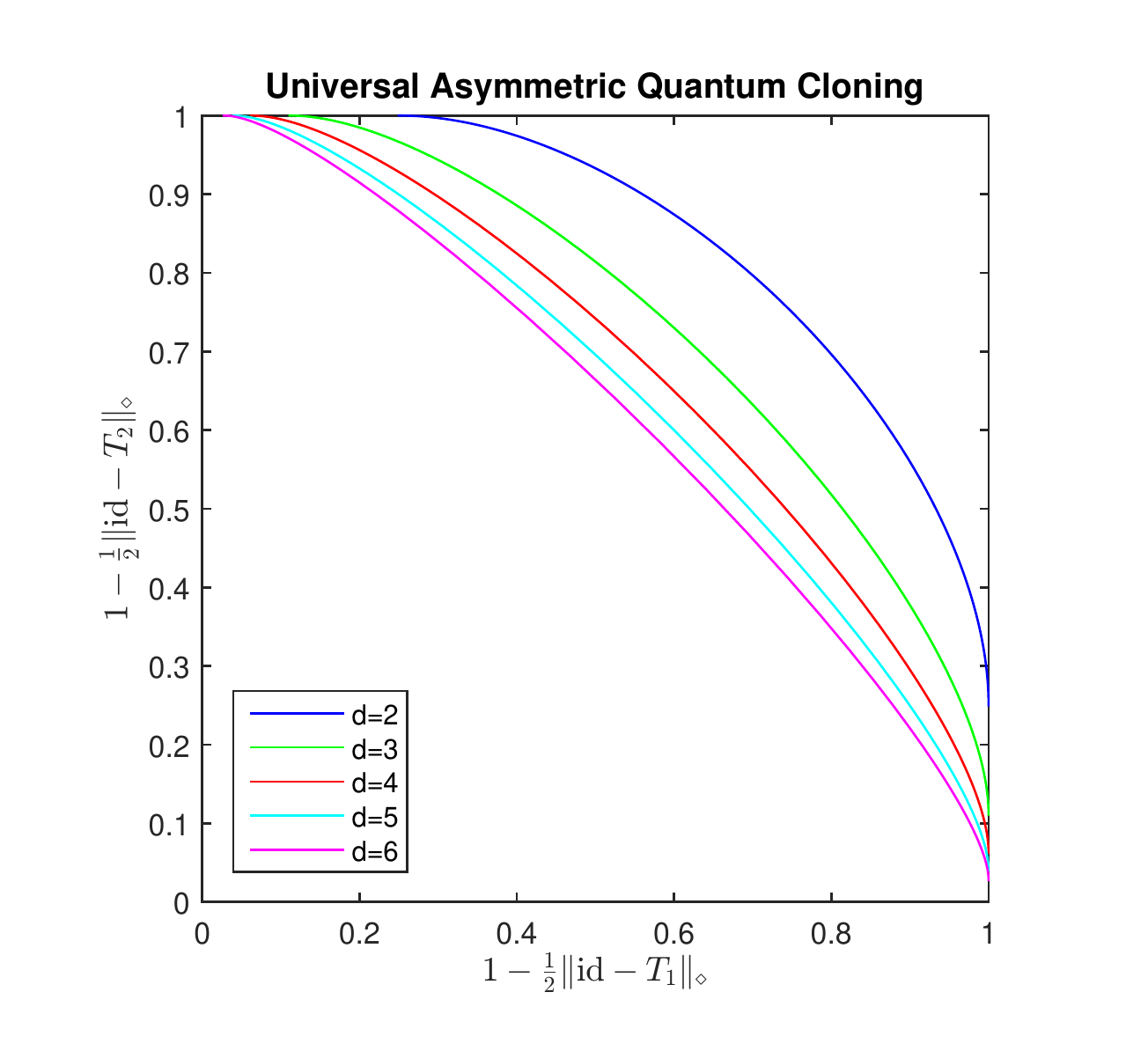}
		\vspace*{13pt}
	\caption[The set of all attainable single quantum clone qualities in terms of $d^\diamond(T_i, \id)=1-\frac{1}{2}\norm{T_i-\id}_\diamond$, $i=1,2$, given by Eq. (\ref{eq:Ck}) for different dimensions of the underlying Hilbert space, using {MATLAB} \cite{Matlab}.]{\label{fig:Picture_ConvexHull_DiamondNorm}The set of all attainable single quantum clone qualities in terms of $d^\diamond(T_i, \id)=1-\frac{1}{2}\norm{T_i-\id}_\diamond$, $i=1,2$, given by Eq.~(\ref{eq:Ck}) for different dimensions of the underlying Hilbert space, using {MATLAB} \cite{Matlab}. The upper boundary of this set is given by
\begin{minipage}{\linewidth}
\begin{equation*}
\frac{1}{d+1} \left( \sqrt{x^{(\diamond)}_1} + \sqrt{x^{(\diamond)}_2} \right)^2  +\frac{1}{d-1} \left( \sqrt{x^{(\diamond)}_1} - \sqrt{x^{(\diamond)}_2} \right)^2 =\frac{2}{d}.
\end{equation*}
\end{minipage}}
	\end{figure}

	\begin{figure}
	\vspace*{13pt}
	\centering
		\includegraphics[trim=110pt 250pt 110pt 250pt, clip, width=1.0\textwidth]{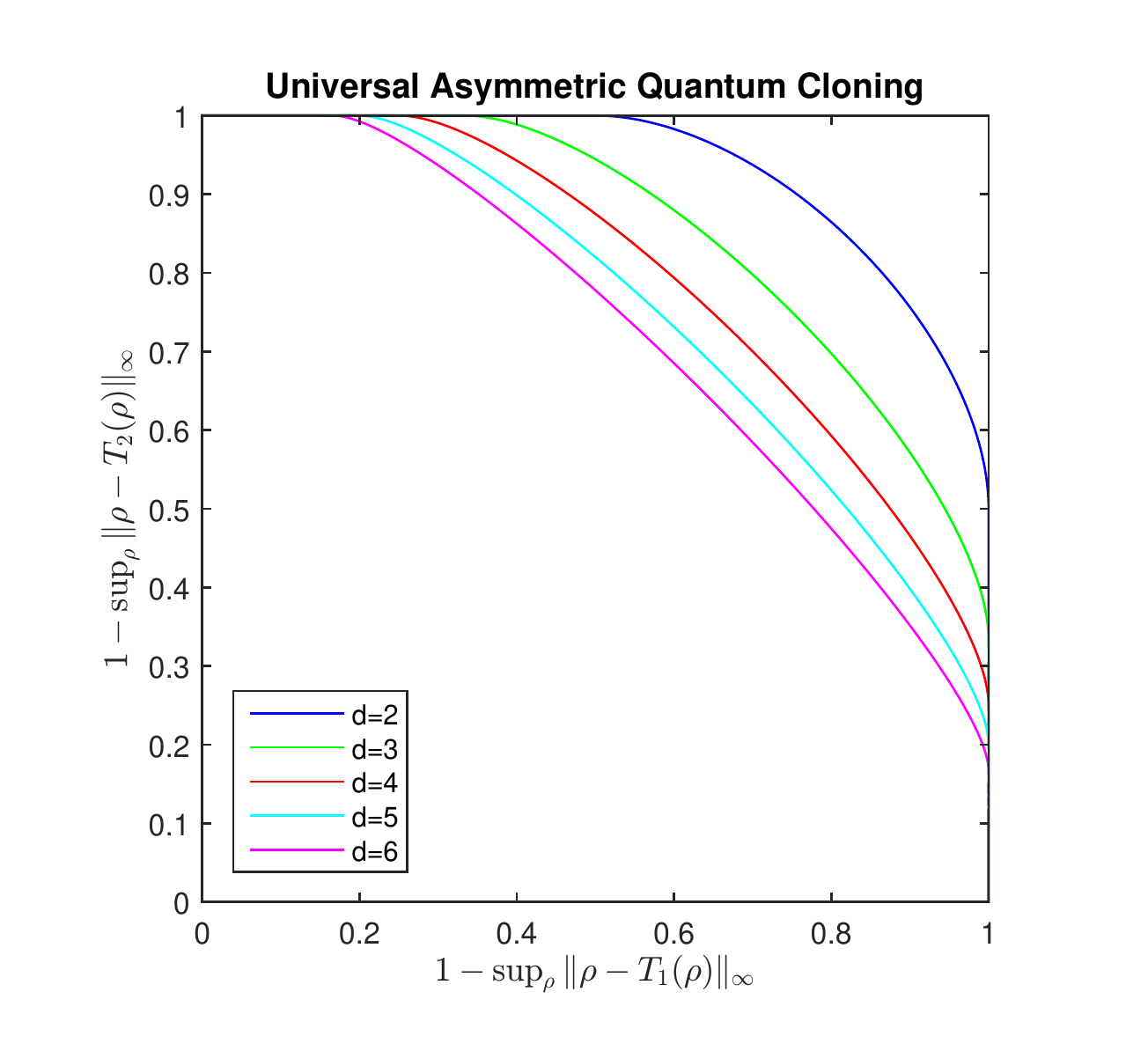}
		\vspace*{13pt}
	\caption[The set of all attainable single quantum clone qualities in terms of $d^\infty(T_i, \id)=1-\sup_\rho \norm{T_i(\rho)-\rho}_\infty$, $i=1,2$, given by Eq. (\ref{eq:Ck}) for different dimensions of the underlying Hilbert space, using {MATLAB} \cite{Matlab}.]{\label{fig:Picture_ConvexHull_OperatorNorm}The set of all attainable single quantum clone qualities in terms of $d^\infty(T_i, \id)=1-\sup_\rho \norm{T_i(\rho)-\rho}_\infty$, $i=1,2$, given by Eq.~(\ref{eq:Ck}) for different dimensions of the underlying Hilbert space, using {MATLAB} \cite{Matlab}. The upper boundary of this set is given by
\begin{minipage}{\linewidth}
\begin{multline*}
\frac{1}{d+1} \left( \sqrt{1+\frac{1+d}{d}\left(x^{(\infty)}_1-1\right)} + \sqrt{1+\frac{1+d}{d}\left(x^{(\infty)}_2-1\right)} \right)^2  \\+\frac{1}{d-1} \left( \sqrt{1+\frac{1+d}{d}\left(x^{(\infty)}_1-1\right)} - \sqrt{1+\frac{1+d}{d}\left(x^{(\infty)}_2-1\right)} \right)^2 =\frac{2}{d}.
\end{multline*}
\end{minipage}}
\end{figure}

\begin{figure}[p]
\vspace*{13pt}
\centering
		\includegraphics[trim=110pt 250pt 110pt 250pt, clip, width=1.0\textwidth]{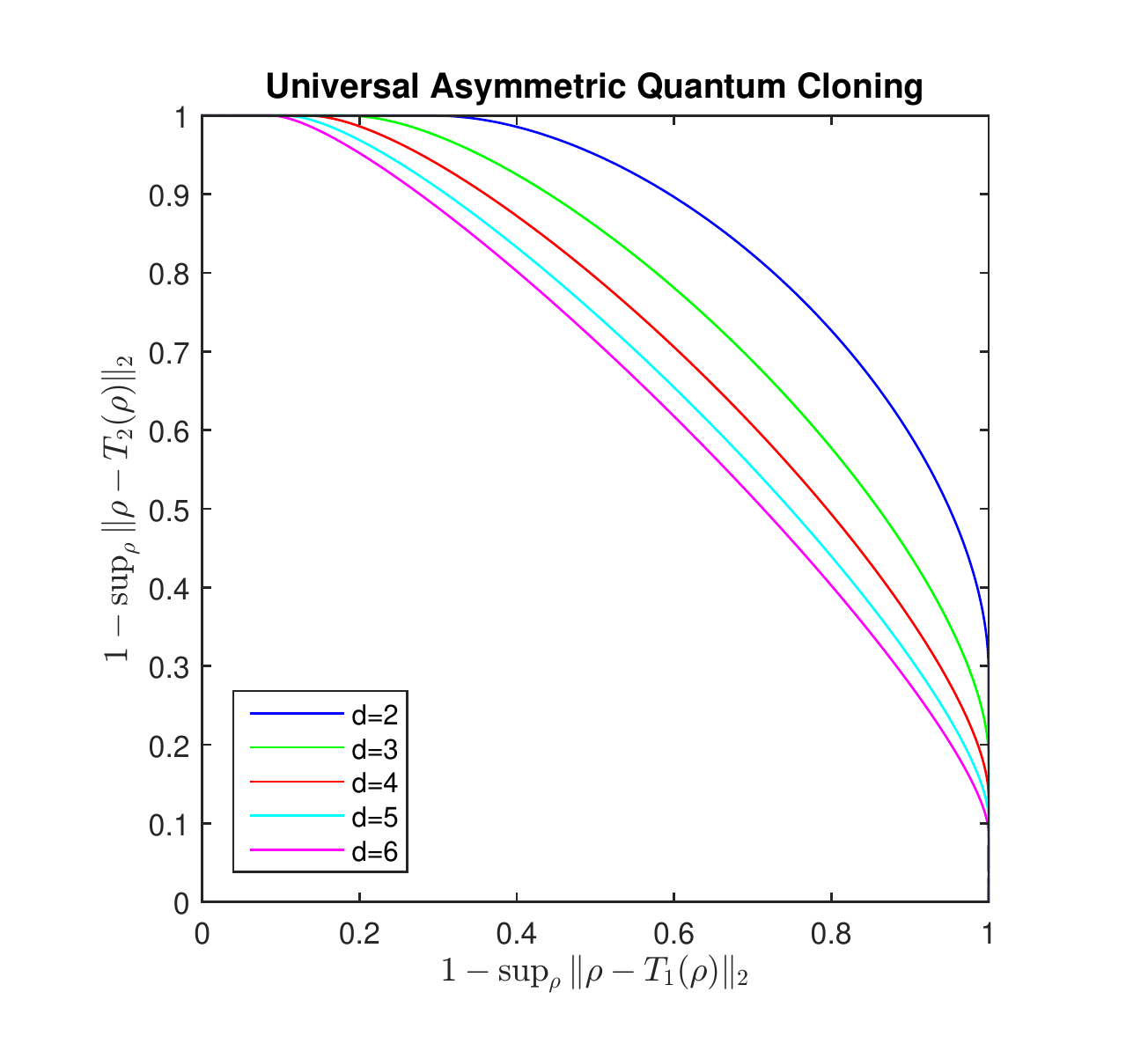}
		\vspace*{13pt}
	\caption[The set of all attainable single quantum clone  qualities in terms of $d^2(T_i, \id)=1-\sup_\rho \norm{T_i(\rho)-\rho}_2$, $i=1,2$, given by Eq. (\ref{eq:Ck}) for different dimensions of the underlying Hilbert space, using {MATLAB} \cite{Matlab}.]{	\label{fig:Picture_ConvexHull_FrobeniusNorm}The set of all attainable single quantum clone  qualities in terms of $d^2(T_i, \id)=1-\sup_\rho \norm{T_i(\rho)-\rho}_2$, $i=1,2$, given by Eq.~(\ref{eq:Ck}) for different dimensions of the underlying Hilbert space, using {MATLAB} \cite{Matlab}. The upper boundary of this set is given by
\begin{minipage}{\linewidth}
\begin{multline*}
\frac{1}{d+1} \left( \sqrt{1+\frac{d^2-1}{d^2}\sqrt{\frac{d}{d-1}}\left(x_1^{(2)}-1\right)} + \sqrt{1+\frac{d^2-1}{d^2}\sqrt{\frac{d}{d-1}}\left(x_2^{(2)}-1\right)} \right)^2 \\ +\frac{1}{d-1} \left( \sqrt{1+\frac{d^2-1}{d^2}\sqrt{\frac{d}{d-1}}\left(x_1^{(2)}-1\right)} - \sqrt{1+\frac{d^2-1}{d^2}\sqrt{\frac{d}{d-1}}\left(x_2^{(2)}-1\right)} \right)^2 =\frac{2}{d}.
\end{multline*}
\end{minipage}}
\end{figure}

\end{document}